\documentclass[
  pre,
  twocolumn,
  twoside,
  byrevtex,
  superscriptaddress,
  floatfix,
  longbibliography,
]{revtex4-2}

\usepackage[realmainfile]{currfile}
\input{\currfilebase.settings}
\usepackage{amssymb}
\usepackage{amsmath}

\usepackage{url}
\PassOptionsToPackage{hyphens}{url}\usepackage{hyperref}
\makeatletter
\g@addto@macro{\UrlBreaks}{\UrlOrds}
\makeatother

\usepackage[table]{xcolor}   
\usepackage{colortbl}

\definecolor{goodblue}{RGB}{0, 91, 187}
\usepackage{hyperref}
\hypersetup{
  colorlinks=true,
  allcolors=goodblue,
  urlcolor=goodblue,
  citecolor=goodblue,
  pdfborder={0 0 0},
  breaklinks=true,
}

\usepackage[normalem]{ulem}

\usepackage{textcomp}

\usepackage[export]{adjustbox}

\makeatletter

\def\CT@@do@color{%
  \global\let\CT@do@color\relax
  \@tempdima\wd\z@
  \advance\@tempdima\@tempdimb
  \advance\@tempdima\@tempdimc
  \advance\@tempdimb\tabcolsep
  \advance\@tempdimc\tabcolsep
  \advance\@tempdima2\tabcolsep
  \kern-\@tempdimb
  \leaders\vrule
  \hskip\@tempdima\@plus  1fill
  \kern-\@tempdimc
  \hskip-\wd\z@ \@plus -1fill }
\makeatother

\usepackage[table]{xcolor}

\definecolor{olivegreen}{rgb}{0.33333,.41961,0.18431}
\definecolor{forestgreen}{rgb}{0.13333,.5451,0.13333}
\definecolor{lightgrey}{rgb}{0.7,0.7,0.7}
\definecolor{verylightgrey}{rgb}{0.90,0.90,0.90}
\definecolor{veryverylightgrey}{rgb}{0.95,0.95,0.95}
\definecolor{grey}{rgb}{0.5,0.5,0.5}

\definecolor{headerblue}{HTML}{33367E}
\definecolor{unitednationsblue}{HTML}{4D88FF}

\definecolor{charcoal}{HTML}{36454F}
\definecolor{cinerous}{HTML}{98817B}
\definecolor{feldgrau}{HTML}{4D5D53}
\definecolor{glaucous}{HTML}{6082B6}
\definecolor{arsenic}{HTML}{3B444B}
\definecolor{xanadu}{HTML}{738678}

\definecolor{firebrick}{HTML}{B22222}
\definecolor{orangered}{HTML}{FF4500}
\definecolor{tomato}{HTML}{FF6347}

\definecolor{purpletaupe}{HTML}{3B444B}

\newcommand{\done}[1]{}

\usepackage{titlesec}
\titleformat*{\paragraph}{\bfseries}

\usepackage{changepage}

\usepackage{graphicx}
\usepackage{epsfig}
\usepackage{verbatim}

\usepackage{enumerate}
\usepackage{enumitem}

\usepackage{amssymb}
\usepackage{amsmath}

\usepackage{ifthen}

\usepackage{longtable}

\usepackage{mathtools}

\newboolean{twocolswitch}

\usepackage{array}

\newcommand{\PreserveBackslash}[1]{\let\temp=\\#1\let\\=\temp}

\newcommand{\sindex}[1]{}
\newcommand{\nindex}[1]{}

\newcommand{\www}[1]{\url{#1}}

\usepackage{lettrine}

\newcommand{\lmat}{\left[
    \begin{array}
    }
    \newcommand{\rmat}{\end{array}
  \right]
}

\setboolean{twocolswitch}{true}

\begin{document}




\title{\protect
  Detecting sub-populations in online health communities:\\ 
A mixed-methods exploration of breastfeeding messages in BabyCenter Birth Clubs

}

\author{
\firstname{Calla}
\surname{Beauregard}
}
\email{calla.beauregard@uvm.edu}

\affiliation{
  Vermont Complex Systems Institute,
  University of Vermont,
  Burlington,
  VT 05405,
  USA.
}

\affiliation{
  Computational Story Lab,
  University of Vermont,
  Burlington,
  VT 05405,
  USA.
}

\author{
\firstname{Parisa}
\surname{Suchdev}
}

\affiliation{
  Vermont Complex Systems Institute,
  University of Vermont,
  Burlington,
  VT 05405,
  USA.
}

\affiliation{
  Computational Story Lab,
  University of Vermont,
  Burlington,
  VT 05405,
  USA.
}

\affiliation{
  Computational Ethics Lab,
  University of Vermont,
  Burlington,
  VT 05405,
  USA.
}

\author{
\firstname{Ashley M. A.}
\surname{Fehr}
}

\affiliation{
  Vermont Complex Systems Institute,
  University of Vermont,
  Burlington,
  VT 05405,
  USA.
}

\affiliation{
  Computational Story Lab,
  University of Vermont,
  Burlington,
  VT 05405,
  USA.
}

\author{
\firstname{Isabelle T.}
\surname{Smith}
}

\affiliation{
  Vermont Complex Systems Institute,
  University of Vermont,
  Burlington,
  VT 05405,
  USA.
}

\author{
\firstname{Tabia}
\surname{Tanzin Prama}
}

\affiliation{
  Vermont Complex Systems Institute,
  University of Vermont,
  Burlington,
  VT 05405,
  USA.
}

\affiliation{
  Computational Story Lab,
  University of Vermont,
  Burlington,
  VT 05405,
  USA.
}

\author{
\firstname{Julia}
\surname{Witte Zimmerman}
}

\affiliation{
  Vermont Complex Systems Institute,
  University of Vermont,
  Burlington,
  VT 05405,
  USA.
}

\affiliation{
  Computational Story Lab,
  University of Vermont,
  Burlington,
  VT 05405,
  USA.
}

\affiliation{
  Computational Ethics Lab,
  University of Vermont,
  Burlington,
  VT 05405,
  USA.
}

\author{
\firstname{Carter}
\surname{Ward}
}

\affiliation{
  Computational Ethics Lab,
  University of Vermont,
  Burlington,
  VT 05405,
  USA.
}

\author{
\firstname{Juniper}
\surname{Lovato}
}

\affiliation{
  Vermont Complex Systems Institute,
  University of Vermont,
  Burlington,
  VT 05405,
  USA.
}

\affiliation{
  Department of Computer Science,
  University of Vermont,
  Burlington,
  VT 05405,
  USA.
}

\affiliation{
  Computational Ethics Lab,
  University of Vermont,
  Burlington,
  VT 05405,
  USA.
}

\author{
\firstname{Christopher M.}
\surname{Danforth}
}

\affiliation{
  Vermont Complex Systems Institute,
  University of Vermont,
  Burlington,
  VT 05405,
  USA.
}

\affiliation{
  Computational Story Lab,
  University of Vermont,
  Burlington,
  VT 05405,
  USA.
}

\affiliation{
  Department of Mathematics and Statistics,
  University of Vermont,
  Burlington,
  VT 05405,
  USA.
}

\author{
\firstname{Peter Sheridan}
\surname{Dodds}
}
\email{peter.dodds@uvm.edu}

\affiliation{
  Vermont Complex Systems Institute,
  University of Vermont,
  Burlington,
  VT 05405,
  USA.
}

\affiliation{
  Computational Story Lab,
  University of Vermont,
  Burlington,
  VT 05405,
  USA.
}

\affiliation{
  Department of Computer Science,
  University of Vermont,
  Burlington,
  VT 05405,
  USA.
}

\affiliation{
  Santa Fe Institute,
  1399 Hyde Park Rd,
  Santa Fe,
  NM 87501,
  USA.
}

\date{\today}

\begin{abstract}
  \protect
  Parental stress is a nationwide health crisis according to the U.S. Surgeon General's 2024 advisory.
To allay stress, expecting parents seek advice and share experiences in a variety of venues, from in-person birth education classes and parenting groups to virtual communities, for example, BabyCenter, a moderated online forum community with over 4 million members in the United States alone. 
In this study, we aim to understand how parents talk about pregnancy, birth, and parenting by analyzing 5.43M posts and comments from the April 2017--January 2024 cohort of 331,843  BabyCenter ``birth club'' users (that is, users who participate in due date forums or ``birth clubs'' based on their babies' due dates). Using BERTopic to locate breastfeeding threads and LDA to summarize themes, we compare documents in breastfeeding threads to all other birth-club content. Analyzing time series of word rank, we find that posts and comments containing anxiety-related terms increased steadily from April 2017 to January 2024. We used an ensemble of topic models to identify dominant breastfeeding topics within birth clubs, and then explored trends among all user content versus those who posted in threads related to breastfeeding topics. We conducted Latent Dirichlet Allocation (LDA) topic modeling to identify the most common topics in the full population, as well as within the subset breastfeeding population. We find that the topic of sleep dominates in content generated by  the breastfeeding population, as well anxiety-related and work/daycare topics that are not predominant in the full BabyCenter birth club dataset. 
\begin{figure}[h] 
    \centering
    \includegraphics[width=0.6\textwidth]{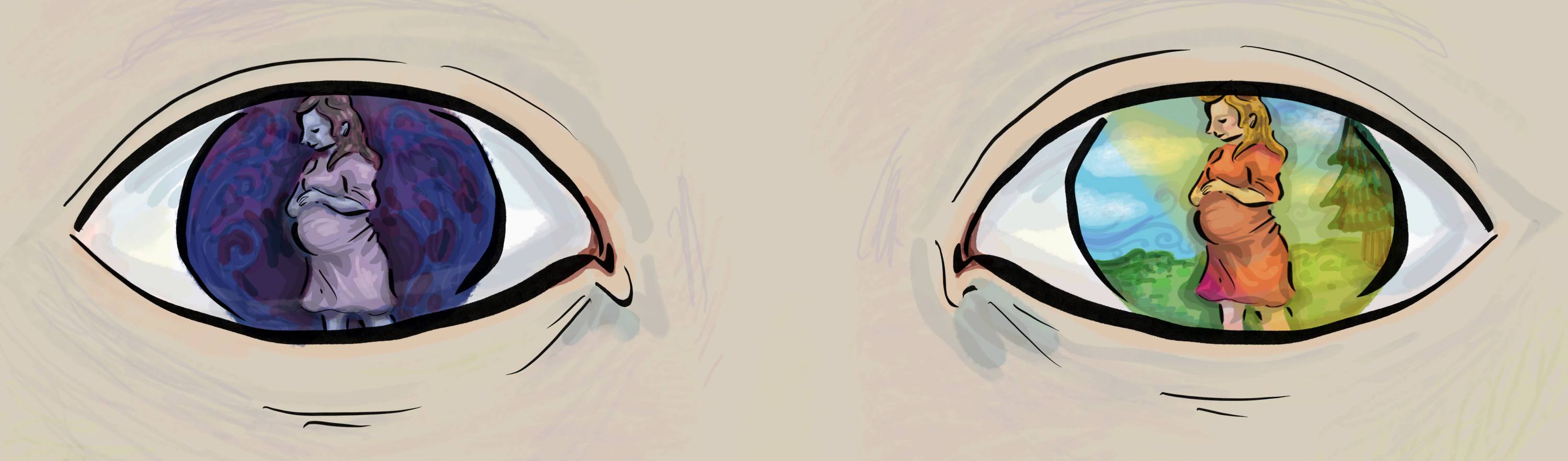} 
    \caption{Figure representing the complex milieu of emotions surrounding fertility, pregnancy, and birth, even within a single person's experience.}
    \label{fig:myimage}
\end{figure}

 
\end{abstract}

\maketitle

\setlength{\parskip}{1\baselineskip plus .1\baselineskip  minus .1\baselineskip}


\section{Introduction}
\label{sec:papertag.introduction}
The Centers for Disease Control (CDC) in the United States has reported that American women are largely dissatisfied with the quality of their maternity care~\cite{cdc_mistreatment_2023}. Based on the largest survey to date, 1 in 5 women report mistreatment during their care. Racial disparity compounds reports of mistreatment, with 1 in 3 Black, Hispanic, and multiracial women reporting mistreatment. Furthermore, 45\% of women report holding back when asking questions of their maternity care providers~\cite{cdc_mistreatment_2023}. Given that demographics (specifically racial identity) can impact endorsement of medical symptoms by providers, at least under some circumstances~\cite{mclean2011gender},\footnote{\citet{mclean2011gender} discusses psychopathology specifically, and since mental health -- like motherhood -- is an intimate, value-laden, culturally-mediated, even stigmatized, health topic, it seems a likely comparison in this regard.} it is plausible that there is interaction between identity and the quality of an individual's patient-provider interaction. 

In addition to poor experience with maternity care, maternal mortality rates are higher in the United States than in comparable nations~\cite{noauthor_maternal_2020}, and particularly high among Black, non-Hispanic women as compared to White women~\cite{hoyert_maternal_2024}.

In order to understand questions of public health, researchers often seek out the `highest quality evidence' in the form of randomized controlled trials, 
but such studies are expensive~\cite{speich_systematic_2018} and have historically excluded women~\cite{liu_womens_2016}. 
Public health research has increasingly focused on big data sources and employed epidemiological techniques to better understand populations~\cite{dolley_big_2018}. Practicing ``digital epidemiology'' (epidemiology on non-traditional data sources, see Salathé's 2018 review~\cite{salathe_digital_2018}) means using publicly available data to identify trends invisible in more traditional formats. 
Social media data, which often chronicles day-to-day mood and activity, has particular promise for mental health research~\cite{stupinski_quantifying_2022}.
However, there is potential for significant bias in individuals' online self-portrayals, as well as ethical questions related to research use of data shared by users who may not have intended it for such purposes~\cite{tusl_opportunities_2022}.

Considering the recent turmoil surrounding women's health funding in the United States and funding for foreign aid programs~\cite{noauthor_hhs_2025, kallen_undermining_2025}, it is vital that public health researchers consider other avenues to understand populations, especially those that historically underserved. Previous work has explored language differences in parenting subreddits between mothers and fathers using topic modeling~\cite{sepahpourfard2022parenting} as well as explored discourse on Mumsnet, a United Kingdom based parenting forum, using an anthropological approach~\cite{locke2021bookreview}. In this work, we employ mixed methods and ensemble topic modeling to identify trends in language associated with users who post about breastfeeding on BabyCenter, a moderated online forum community with over 4 million members in the United States alone. 
We specifically examine breastfeeding due to its important relationship with maternal mental health. Anxiety disorders are more common in postpartum individuals than in the general population, treatment rates are low for postpartum anxiety, and the experience of postpartum anxiety can be associated with early breastfeeding cessation~\cite{ali2018womens}. Accordingly, recent research demonstrates a significant relationship between the duration of breastfeeding and subjective norms~\cite{davies_exclusive_2022}, guilt and poor maternal mental health~\cite{jackson_guilt_2021}, and guilt and the shorter duration of exclusive breastfeeding~\cite{russell_infant_2021}. A cross-sectional online survey ($N=470$) of mothers 6- and 12-month postpartum suggests that ``postpartum anxiety may be an underlying mechanism which reduces exclusive breastfeeding duration and negatively affects maternal perceptions of infant sleep quality''~\cite{davies_exclusive_2022}. A lack of studies on postpartum mental health~\cite{ali2018womens}, and evidence that anxiety disorders are more prevalent and burdensome in women than in men further motivates the study of the impact and manifestation of anxiety with respect to other aspects of women's health~\cite{mclean2011gender}.\footnote{About one in three women will meet the criteria for anxiety disorders during their lifetime, versus roughly one in five men~\cite{mclean2011gender}.}

First, we conduct topic modeling using the \texttt{BERTopic} Python package~\cite{grootendorst_bertopic_2022} to understand the nature of posts and comments in the April 2019 birth club, the largest such club on BabyCenter. 
We identify the most prominent topics related to breastfeeding across all clubs. We then use Latent Dirichlet Allocation (LDA) to examine trends in language use across both the entire dataset (April 2017--January 2024) and the subset of users who post in breastfeeding topics. 
We conclude with an analysis of topics inferred from word rank time series and allotaxonometry~\cite{dodds_allotaxonometry_2023}.

\section{Description of datasets}
\label{sec:papertag.data}

The present study uses posts and comments from all US BabyCenter monthly birth clubs spanning April 2017--January 2024 ($n=5,433,338$ posts and comments by $n=331,843$ unique users).
Each birth club represents a cohort of US-based BabyCenter users with expected due dates in that club's given month and year. Users post the majority of their content in the months leading up to the delivery date (their club's namesake), with activity sometimes continuing but generally tapering off after delivery. 

\begin{figure}[htbp]
    \centering
    \includegraphics[width=0.42\textwidth]{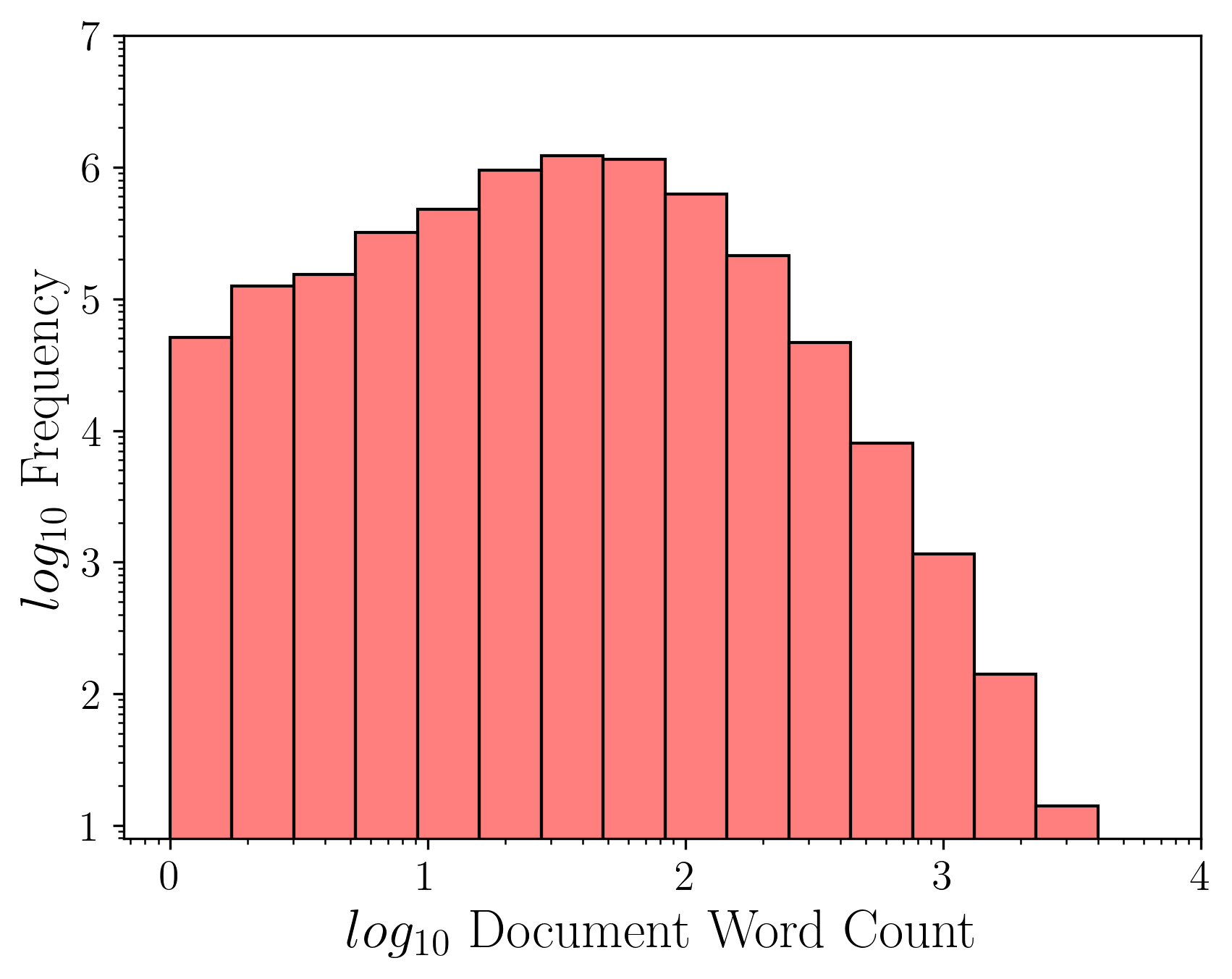}
    \caption{\textbf{Histogram of Words of All Documents using kernel density estimation (KDE) for smoothing.} The most common documents (posts and comments), appearing approximately 1 million times in the dataset, are roughly 30-40 words each, displayed by the bin height.}
    \label{fig:histogram_alldocs}
\end{figure}

The dataset contains a unique username identifier and the calendar date for each post or comment. For clarity, we examine all posts and comments and will refer to them as ``documents'' for the remainder of this paper. Figure~\ref{fig:histogram_alldocs} shows the distribution of word counts across all documents in the dataset. 

\begin{figure*}[htbp]
    \centering
    \includegraphics[width=1\textwidth]{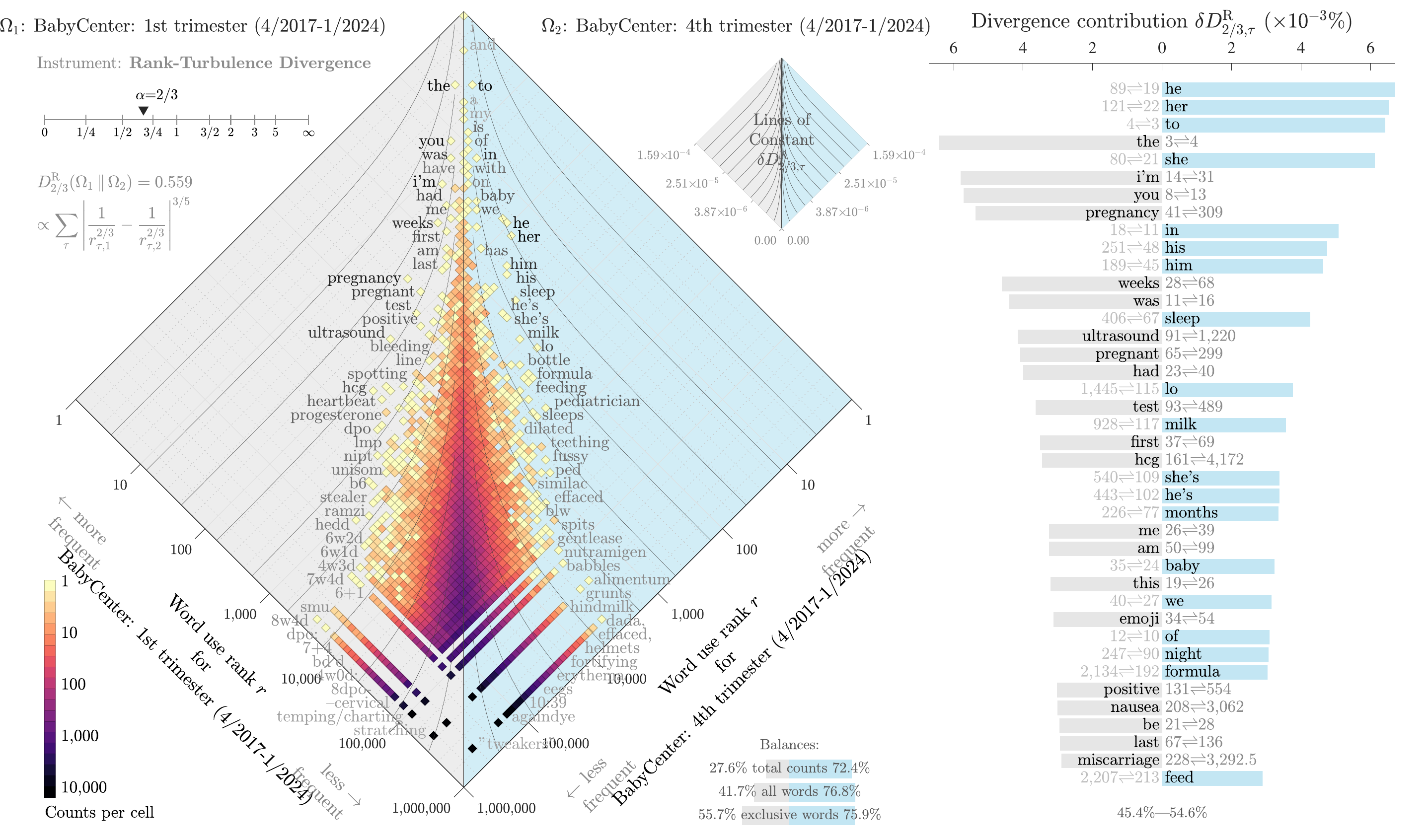}
    \caption{\textbf{Allotaxonometric Representation of First and Fourth Trimesters, April 2017--January 2024}. In the Allotaxonograph, the first trimester appears on the left-hand panel, and the fourth trimester appears on the right-hand panel. Reading from top to bottom, the highest-ranked words in each corpus appear sequentially in descending order.}
    \label{fig:allotax}
\end{figure*}

To validate longitudinal temporal trends in user posts and comments in birth clubs, we first use allotaxonometry (the study of rank turbulence, see Dodds et al. 2023~\cite{dodds_allotaxonometry_2023}). We combine data from every birth club based on trimester thresholds using the first day of the due date month, and partition the data by three-month increments. We create groups from this data transformation to represent all posts in the first trimester and all posts in the fourth trimester. We then compare these groups using the Allotaxonometer. In Figure~\ref{fig:allotax}, we observe that words like ``ultrasound'' appear more frequently by rank in the first trimester but pronouns like ``her'', ``she'' and ``him'' as well as the word ``baby'' occur more frequently in the fourth trimester. Accordingly, we consider birth clubs as ``cohorts'' for this study, and assume that each club contains longitudinal discussion of pregnancy and birthing. 

\section{Methods}

\subsection{Identifying sub-populations with ensemble topic modeling}
We use an ensemble of topic modeling, starting with BERTopic, to find longitudinal themes in posts and comments across birth club cohorts, specifically in populations discussing breastfeeding. According to its developers, ``by default, the main steps for topic modeling with BERTopic are sentence-transformers, UMAP, HDBSCAN, and c-TF-IDF run in sequence''~\cite{grootendorst_bertopic_2022}. Essentially, sentence transformers embed textual data in matrices, and UMAP (or Uniform Manifold Approximation) performs dimensional reduction of the high-dimensional embedding matrices~\cite{mcinnes_umap_2020}. HDBSCAN (or Hierarchical Density-Based Spatial Clustering of Applications with Noise) determines high-density clusters embedding~\cite{mcinnes_hdbscan_2017}, and the c-TF-IDF (Class-based Term Frequency-Inverse Document Frequency) returns terms and topics based on relative presence in documents~\cite{grootendorst_bertopic_2022}. We retain the order of these steps for this analysis and use a random seed of $42$ in order to replicate the stochastic aspects of topic modeling. We then identify the top breastfeeding topic in each birth club through keyword search of topic names and representative documents in each topic. We are highly inclusive in our selection and ensure that we include terms like ``ebf'' (i.e., exclusive breastfeeding) and ``pumping'' as alternate breastfeeding terms. We aim to select the largest breastfeeding topic so long as it was topically related to breastfeeding experience. For example, some breastfeeding topics are primarily about ``nursing bras''. In this case, we select the next largest, topically related breastfeeding topic, e.g., ``infant feeding''. 

We first analyze each birth club using topic analysis and then qualitatively code documents within topics to further sample the population using three human coders. Employing this labeled dataset, we test a variety of common machine learning classifiers using a random $80/20$ train/test split of the annotated data. We found that the classifiers did not perform well, even when taking the union of agreement between the three best classifiers and conducting human annotation. Essentially, the classifiers disagreed on different posts and comments with little to no interpretability, despite reasonably high accuracy scores. For completeness, we report our qualitative coding scheme and accuracy scores in the Appendix (see Table~\ref{table:classifiers}). Based on this lack of interpretability, we employed a simpler approach and subset the population by all users who posted in the largest breastfeeding topic in each birth club. We include all documents in all topics in the particular birth club for each user who posted in the breastfeeding topic.


\subsection{Analysis of topics and breastfeeding sub-population}

We adopt an ensemble approach to topic modeling, combining the strengths of multiple algorithms to improve classification accuracy and interpretability. This method follows prior work demonstrating the effectiveness of topic modeling ensembles~\cite{george_integrated_2023}. For the initial document labeling, we employ BERTopic, which leverages contextual embeddings and clustering techniques. This method is particularly suitable for natural language processing (NLP) tasks requiring a nuanced understanding of document semantics.

We then apply Latent Dirichlet Allocation (LDA), a widely used generative probabilistic model that discovers latent topics in discrete data sets such as text corpora~\cite{kherwa2020topic, egger2022topic}. The primary goal of topic modeling is to uncover the main themes present in unstructured textual data~\cite{rogers2017comparison}. LDA groups words based on semantic similarity~\cite{alkhafaji2021topic}. Each word within a topic is associated with a conditional probability, indicating its relevance to the respective topic cluster. The resulting word clusters represent distinct themes within the corpus. A limitation of the LDA model is that it does not inherently assign descriptive labels to the topics it generates~\cite{Rahman2020ModelingTS}. To address this, we manually label each topic by interpreting its top keywords. We perform a sweep across the best number of topics ($k$) and two distributional parameters ($\alpha$ and $\eta$) to determine the hyperparameters that statistically yield the best goodness of fit via perplexity~\cite{benites-lazaro_topic_2018,egger2022topic,latent_dirichlet} (refer to Appendix Table \ref{parameter_sweep} for all perplexity score results). In our implementation, we extract the top 30 keywords across 20 distinct topics based on the lowest perplexity score across the entire dataset. To ensure accurate labeling, domain experts review and interpret the keywords associated with each topic, which remains the most effective method for interpreting unlabeled topic model outputs. Subsequently, we construct topic-document matrices, where each entry represents the probability of a document belonging to a specific topic. These matrices facilitate the identification of representative keywords for each topic and their association with individual documents. We use these same parameter selections for both the entire dataset, and the subset for consistency. 

\section{Results}
\label{sec:papertag.results}

\begin{figure}[htbp]
    \centering
    \includegraphics[width=0.49\textwidth]{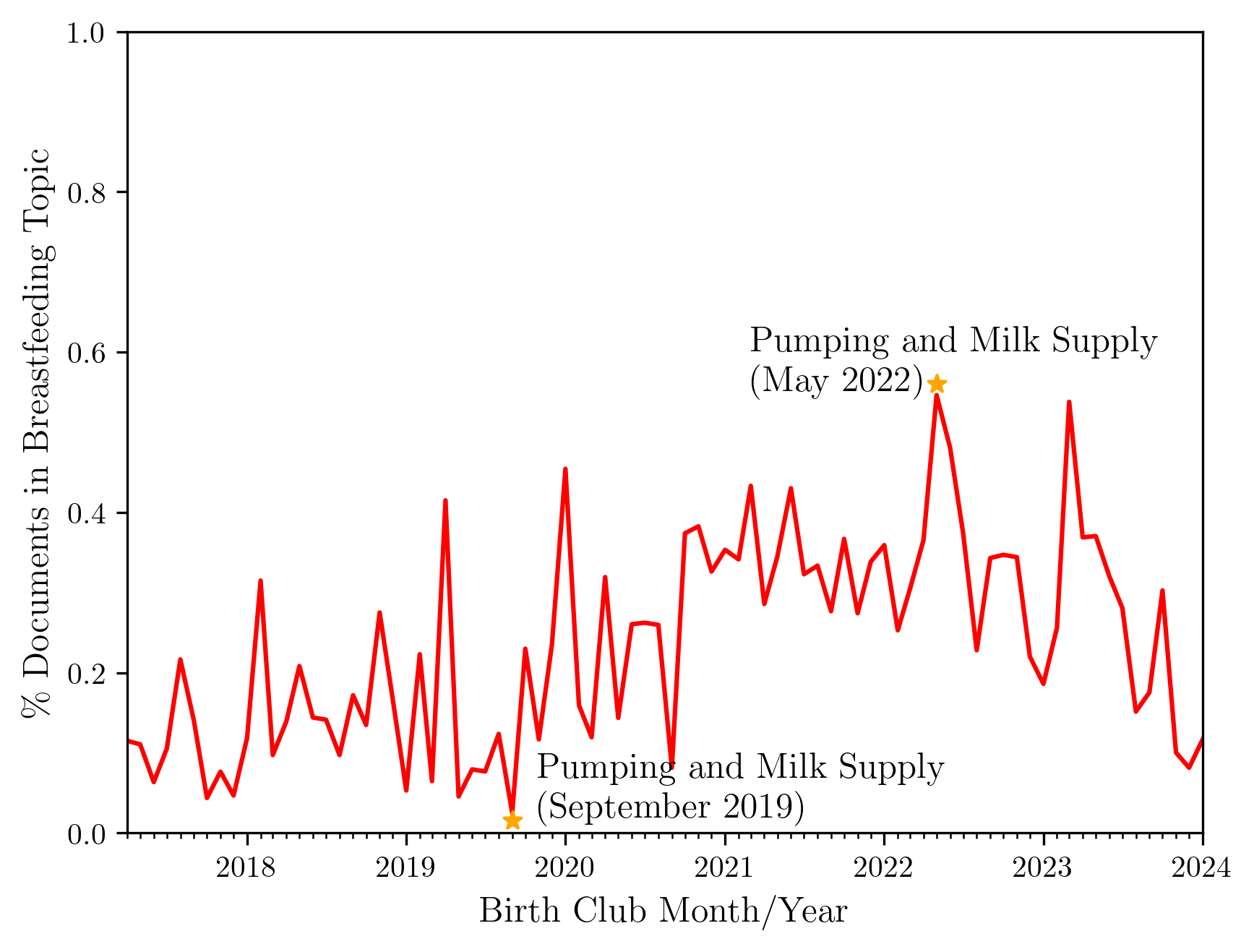}
    \caption{\textbf{Time Series of Percentage of Total Documents for Dominant Breastfeeding Topics, Birth Clubs April 2017--January 2024.} The proportion of breastfeeding topics in monthly birth clubs vacillates over time. The birth clubs with the largest (May 2022) and smallest (September 2019) breastfeeding topics by percent share are marked with a star.}
    \label{fig:timeseries_inclusion}
\end{figure}

\begin{figure}[bp]
    \centering
    \includegraphics[width=0.42\textwidth]{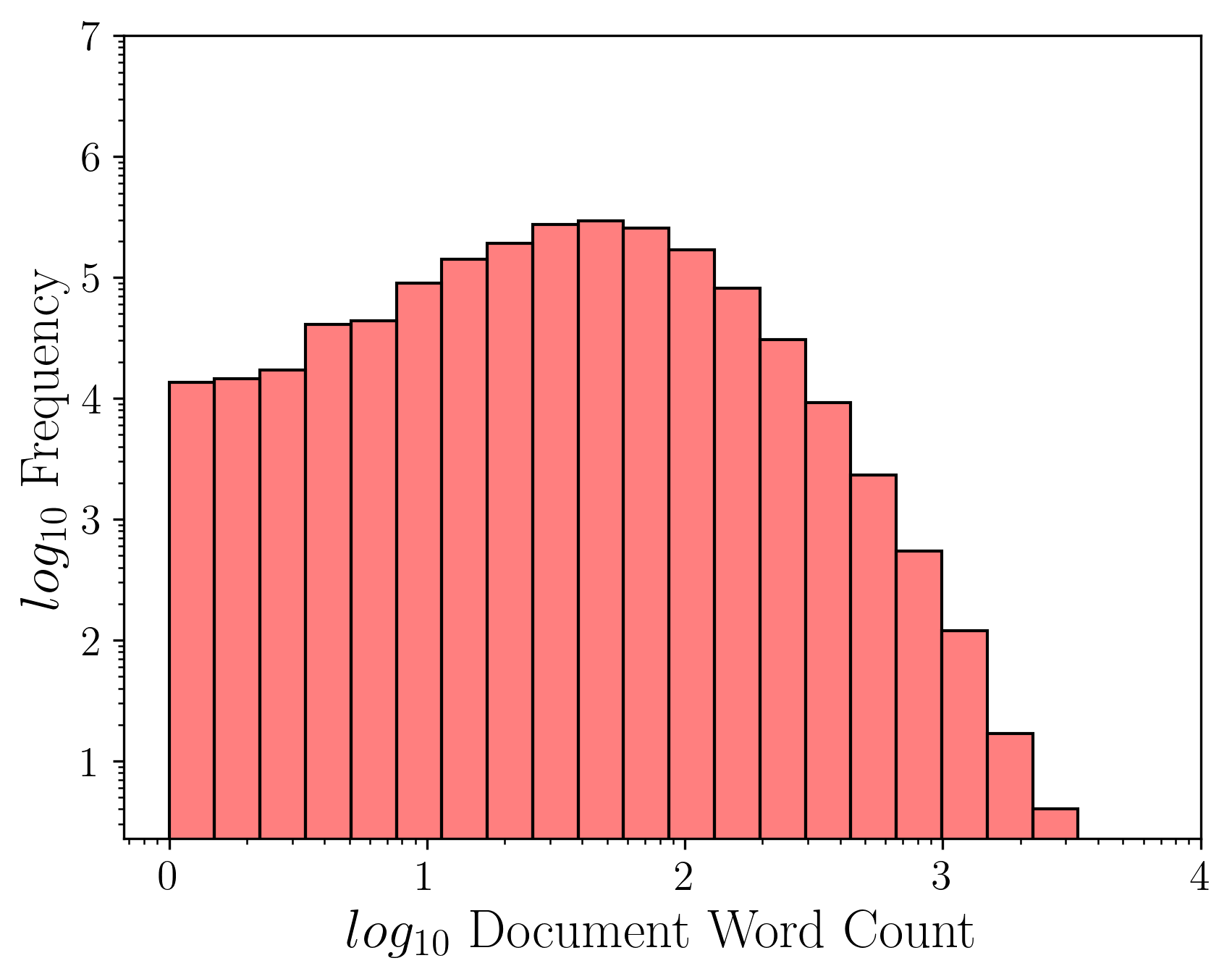}
    \caption{\textbf{Histogram of Words of Breastfeeding Documents using Kernel Density Estimation (KDE) for Smoothing}. The most common documents (posts and comments), appearing approximately 300,000 times in the breastfeeding subset, are roughly 30-40 words each, displayed by the bin height. This shape is very similar to the distribution for the total birth club dataset.}
    \label{fig:histogram_bf}
\end{figure}

In Figure~\ref{fig:timeseries_inclusion}, both the largest and smallest topics by birth club are general breastfeeding content. We manually verify the largest topics related to breastfeeding by inspecting representative documents to limit the inclusion of tangentially related documents. We note that there is still considerable variation across topics in birth clubs due to the stochasticity inherent in UMAP clustering and other features of topic modeling. Unfortunately, this limitation is only verifiable upon manually inspecting every document. In Figure~\ref{fig:histogram_bf}, we further observe that the distribution of word frequencies across the subset of breastfeeding documents has a similar shape to the distribution of word frequencies by document in the full corpus. This suggests that our subset of breastfeeding posts and comments does not deviate substantially in content length from the full dataset. 

\subsection{Trends in word use in full population and breastfeeding sub-population}

\begin{figure*}[htbp]
    \centering
    \includegraphics[width=0.6\textwidth]{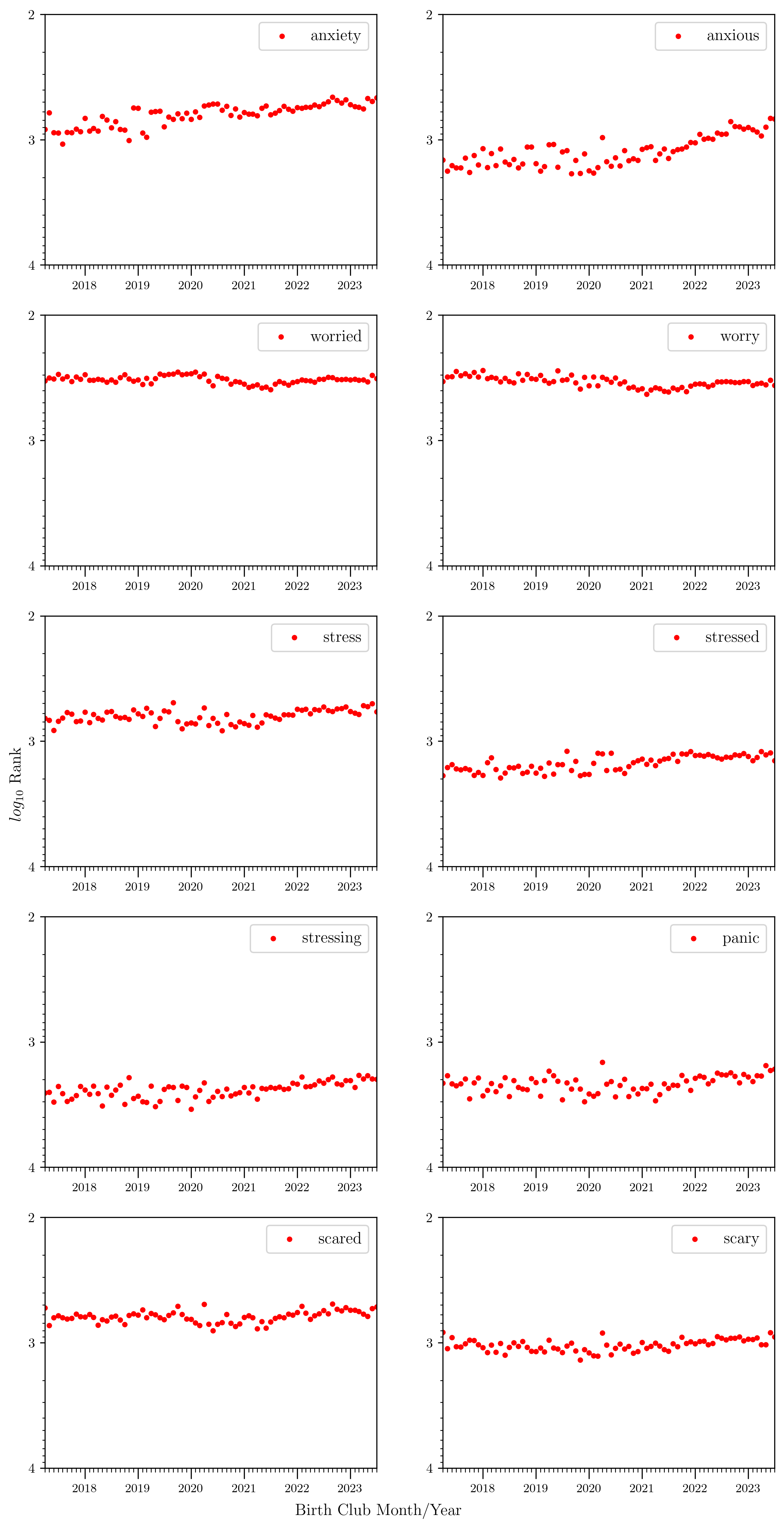}
    \caption{\textbf{Rank Time Series of Diagnostic Statistical Manual (DSM) terms for ``anxiety'' and similar terms, BabyCenter birth clubs (April 2017--June 2023).} Anxiety terms increase over time across birth clubs, with the terms ``anxiety'' and ``anxious'' showing the greatest increases over magnitudes.)
}
    \label{fig:anxiety_over_time}
\end{figure*}

From April 2017--June 2023, terms related to anxiety consistently increase across orders of magnitude, demonstrated in Figure~\ref{fig:anxiety_over_time}. We exclude July 2023--January 2024 birth clubs in the plot to ameliorate any truncated birth clubs that have limited follow-up, as the dataset ends in January 2024. We derived these words using synonyms for anxiety from the Diagnostic Statistical Manual 5th edition (DSM-V)~\cite{apa_dsm5tr_2022}. 

\subsection{Trends in topics in full population and breastfeeding sub-population}

We perform LDA on the full dataset and compare its topics to those of the sub-population. For the full dataset, posts and comments in the ``Work/Family'' topic dominate proportionally across April 2017--June 2023 birth clubs. A ``Baby Sleep'' topic ranks 9th in the full dataset, while it ranks 4th in the breastfeeding subset. Additionally, while ``anxiety'' does not emerge as a topic in the full dataset, the ``anxiety''-related topic in the sub-population follows the same increasing importance as word rank. We also notice a ``Family/work/leave'' topic in the breastfeeding  group that is not present in the total dataset group. Considering the importance of workplace accommodation in breastfeeding and pumping after returning to work, users participating in breastfeeding threads discuss work/leave more than all users. 

\begin{figure*}[htbp]
    \centering
    \includegraphics[width=\textwidth]{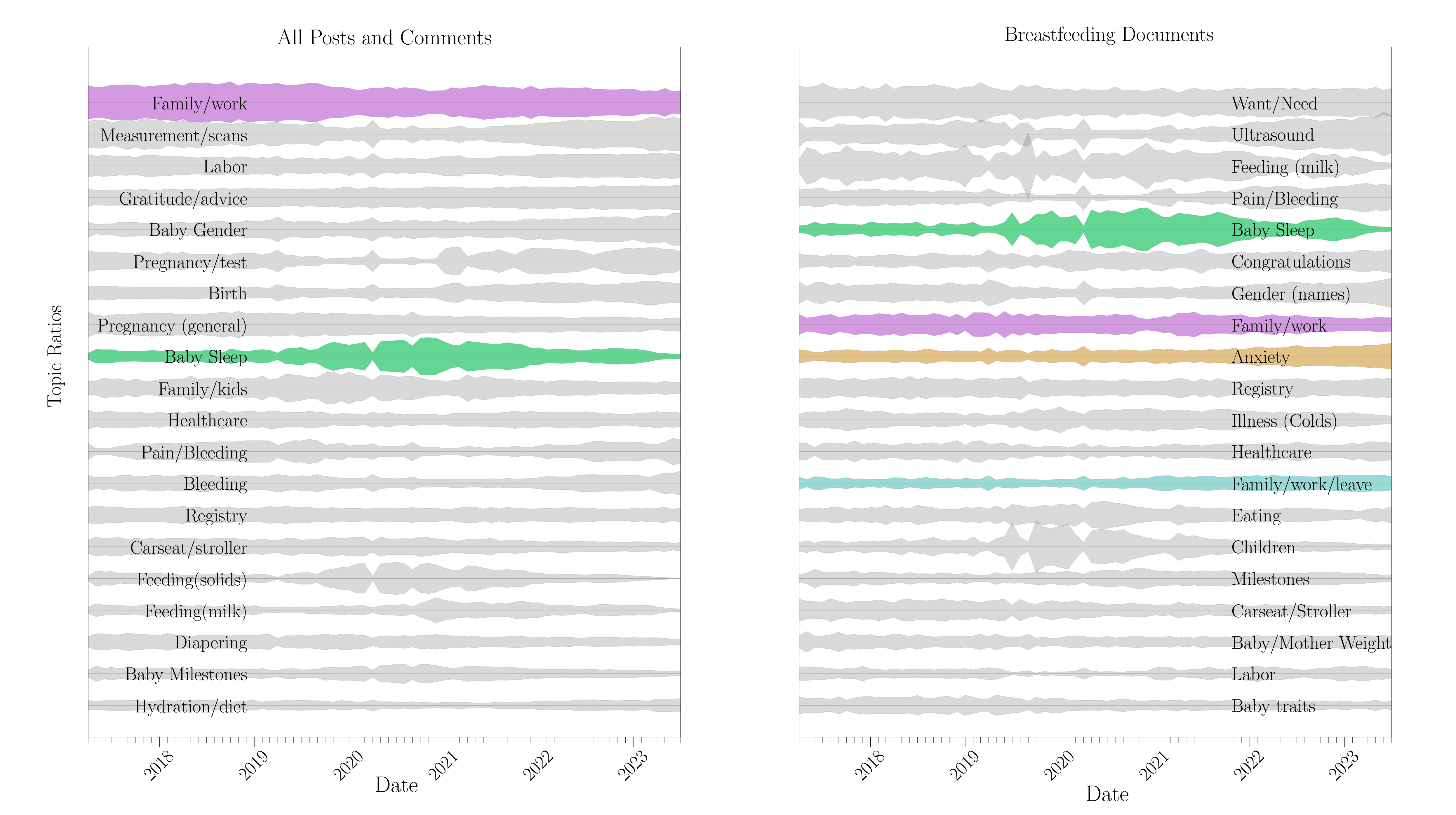}
    \caption{\textbf{Normalized Topic Proportions by BabyCenter Birth Club}. These panels represent the topic proportions ranked by size of the top 20 topics in the full birth club dataset (left) and of all posts and comments by users who posted in dominant breastfeeding topics (right). Shared topics across both datasets have the same color. While ``Family/Work'' appears across both groups, ``anxiety'' only appears as a dominant topic for the breastfeeding subpopulation.}
    \label{fig:sidebyside}
\end{figure*}

\section{Discussion} 
\label{sec:papertag.discussion}
Our mixed topic modeling scheme identifies a subset of documents related to breastfeeding, generating a novel dataset from which to conduct more traditional public health research. A non-exhaustive review of these documents reveals a breadth of sentiment and experience with breastfeeding that is reflected across other forms of research~\cite{ystrom_breastfeeding_2012, coo_role_2020, nagel_maternal_2022}. Recent research has associated non-exclusive breastfeeding (that is, breastfeeding supplemented with formula) to higher levels of anxiety and depression in a study of 229 women who were followed pre-birth to 3--6 months postpartum~\cite{coo_role_2020}. Topic headings generated by topic modeling (either LDA or BERTopic) can obfuscate important information contained in posts/comments, making it difficult to correctly classify nuanced text reflecting specific shame or embarrassment related to breastfeeding~\cite{thomson_shame_2014}. However, the absence of \textit{any} ``anxiety'' or ``leave'' topics in the full dataset topics strongly suggests that users who post in the breastfeeding topics comment more readily on anxiety, as they may experience more anxiety compared to the entire population posting in birth clubs. Furthermore, the specific inclusion of ``leave'' in the breastfeeding documents suggests the importance of workplace accommodation in breastfeeding and pumping after returning to work. 

Based on the increasing understanding of the interplay between mental health and breastfeeding, the relative prominence of sleep-related topics in the breastfeeding sub-population versus the general population is also worth exploring further. The positive impact of adequate sleep on mental health has been well-established~\cite{scott_improving_2021}, which may explain why sleep discussions are prevalent in a group that self-reports many anxious feelings. However, without establishing causality, it is also possible that poor sleep may have contributed to breastfeeding struggles, and sleep improvement interventions may be valuable tools to increase the likelihood of continued breastfeeding. Likewise, the prominence of leave-related discussions in the breastfeeding struggles sub-population may suggest interventions that improve workplace accommodation for breastfeeding may be worth prioritizing as a target for improvement~\cite{bai_lactation_2013,magner_using_2015,tsai_impact_2013}. 

More globally, our results demonstrate a method for examining discussion in online communities for use in public health research. Using topic modeling, distant reading of online communities with NLP can supplement and enhance survey data. Additionally, since this data arises from a self-selected internet community, it is possible that it is more self-reflective and data-rich, as indicated by studies on internet surveys~\cite{hanna_internet-based_2005}. Furthermore, traditional surveys usually occur at a single point in time, in a specific setting (e.g., a screening form in a doctor's office), while our research considers a longer context window. Our approach considered longitudinal information across pregnancy and the postpartum period, thus providing a larger context window regarding birthing and parenting experiences, with our results suggesting important concerns for expecting and new parents that one-time measures may miss. Expanding the context window longitudinally can further seed topics of concern to ask about in surveys or doctor office intake forms. Additionally, surveys and controlled studies often focus on subpopulations of motivated participants who are convenience-sampled at the study's location, particularly when studies have in-person or clinical components, which can introduce a source of bias. The methods we follow to curate a sub-population allow opt-in of many concerned parties that may otherwise not have access to these studies. 

However, it is important to underscore some of the limitations of this study. First, using \texttt{BERTopic} depends on a pipeline of distinct algorithms, which introduces reproducibility issues for future researchers. For instance, as evidenced by the alternative topic models provided in the Appendix, UMAP yields different clusters in each instance of topic analysis and how we decide to represent them can lead to varied interpretations. Secondly, self-disclosed information collected in a non-research environment is not subject to the controls of well-designed research studies and should be closely examined for bias and missing data. 

\section{Concluding remarks}
\label{sec:papertag.concludingremarks}
Considering the limitations of traditional surveys (measurement, coverage, and response bias)~\cite{andrade_limitations_2020, coughlan_survey_2009}, as well as reluctance by birthing people to relate issues to their providers~\cite{cdc_mistreatment_2023}, exploring online communities for risk and protective factors for various conditions is an effective time- and cost-saving venture. Our work makes a first step in identifying an important sub-population of breastfeeding people, employing an ensemble of NLP and topic modeling techniques to more expansively study concerns in this sub-population.  
\section{Ethics Statement}
This study analyzed publicly available data from the US BabyCenter website, a large online health and parenting community. In complaince with our institutional guideline the project was reviewed using the university’s human subjects determination tool and was determined not to constitute human subjects research. Therefore, it did not require formal Institutional Review Board (IRB) review or approval. All analyses complied with BabyCenter’s Terms of Service and respect for user privacy. Posts were accessed in aggregate, and no attempts were made to contact users or link usernames across platforms. To further preserve anonymity, no quoted text is revealed to remove potentially identifying details, and no direct user handles are reported.

\section{Data Availability}
The data used in this study were derived from publicly accessible discussion forums hosted on BabyCenter (\url{www.babycenter.com}) and were collected in compliance with the site’s Terms of Service. Because the raw text of posts may contain potentially identifying or sensitive personal information, we do not make raw text data publicly available. Researchers interested in reproducing analytic steps may request derived, de-identified metadata (e.g., topic assignments, document counts, or aggregated term frequencies) and analysis code from the corresponding author upon reasonable request.

\hspace{1cm}
\acknowledgements
We are grateful for conversations with Bradford Demarest on Topic Analysis. Additionally, we thank Carter Ward, Sarah Nowak, and Jay Lobell for their management of the data from which this was built.

Our work was supported in part by 
MassMutual, 
National Science Foundation Award \#242829 
(Science of Online Corpora, Knowledge, and Stories), 
and a philanthropic gift from an anonymous source. Any opinions, findings, conclusions, or recommendations expressed in this material are those of the authors and do not necessarily reflect the views of the aforementioned supporters. 


\bibliography{\filenamebase.bib}

\begin{thebibliography}{45}%
\makeatletter
\providecommand \@ifxundefined [1]{%
 \@ifx{#1\undefined}
}%
\providecommand \@ifnum [1]{%
 \ifnum #1\expandafter \@firstoftwo
 \else \expandafter \@secondoftwo
 \fi
}%
\providecommand \@ifx [1]{%
 \ifx #1\expandafter \@firstoftwo
 \else \expandafter \@secondoftwo
 \fi
}%
\providecommand \natexlab [1]{#1}%
\providecommand \enquote  [1]{``#1''}%
\providecommand \bibnamefont  [1]{#1}%
\providecommand \bibfnamefont [1]{#1}%
\providecommand \citenamefont [1]{#1}%
\providecommand \href@noop [0]{\@secondoftwo}%
\providecommand \href [0]{\begingroup \@sanitize@url \@href}%
\providecommand \@href[1]{\@@startlink{#1}\@@href}%
\providecommand \@@href[1]{\endgroup#1\@@endlink}%
\providecommand \@sanitize@url [0]{\catcode `\\12\catcode `\$12\catcode `\&12\catcode `\#12\catcode `\^12\catcode `\_12\catcode `\%12\relax}%
\providecommand \@@startlink[1]{}%
\providecommand \@@endlink[0]{}%
\providecommand \url  [0]{\begingroup\@sanitize@url \@url }%
\providecommand \@url [1]{\endgroup\@href {#1}{\urlprefix }}%
\providecommand \urlprefix  [0]{URL }%
\providecommand \Eprint [0]{\href }%
\providecommand \doibase [0]{https://doi.org/}%
\providecommand \selectlanguage [0]{\@gobble}%
\providecommand \bibinfo  [0]{\@secondoftwo}%
\providecommand \bibfield  [0]{\@secondoftwo}%
\providecommand \translation [1]{[#1]}%
\providecommand \BibitemOpen [0]{}%
\providecommand \bibitemStop [0]{}%
\providecommand \bibitemNoStop [0]{.\EOS\space}%
\providecommand \EOS [0]{\spacefactor3000\relax}%
\providecommand \BibitemShut  [1]{\csname bibitem#1\endcsname}%
\let\auto@bib@innerbib\@empty
\bibitem [{\citenamefont {for Disease~Control}\ and\ \citenamefont {Prevention}(2023)}]{cdc_mistreatment_2023}%
  \BibitemOpen
  \bibfield  {author} {\bibinfo {author} {\bibfnamefont {C.}~\bibnamefont {for Disease~Control}}\ and\ \bibinfo {author} {\bibnamefont {Prevention}},\ }\href {https://www.cdc.gov/vitalsigns/respectful-maternity-care/index.html} {\bibinfo {title} {Mistreatment during maternity care}} (\bibinfo {year} {2023})\BibitemShut {NoStop}%
\bibitem [{\citenamefont {McLean}\ \emph {et~al.}(2011)\citenamefont {McLean}, \citenamefont {Asnaani}, \citenamefont {Litz},\ and\ \citenamefont {Hofmann}}]{mclean2011gender}%
  \BibitemOpen
  \bibfield  {author} {\bibinfo {author} {\bibfnamefont {C.~P.}\ \bibnamefont {McLean}}, \bibinfo {author} {\bibfnamefont {A.}~\bibnamefont {Asnaani}}, \bibinfo {author} {\bibfnamefont {B.~T.}\ \bibnamefont {Litz}},\ and\ \bibinfo {author} {\bibfnamefont {S.~G.}\ \bibnamefont {Hofmann}},\ }\bibfield  {title} {\bibinfo {title} {Gender differences in anxiety disorders: Prevalence, course of illness, comorbidity and burden of illness},\ }\href {https://doi.org/10.1016/j.jpsychires.2011.03.006} {\bibfield  {journal} {\bibinfo  {journal} {Journal of Psychiatric Research}\ }\textbf {\bibinfo {volume} {45}},\ \bibinfo {pages} {1027} (\bibinfo {year} {2011})}\BibitemShut {NoStop}%
\bibitem [{Note1()}]{Note1}%
  \BibitemOpen
  \bibinfo {note} {\protect \citet {mclean2011gender} discusses psychopathology specifically, and since mental health -- like motherhood -- is an intimate, value-laden, culturally-mediated, even stigmatized, health topic, it seems a likely comparison in this regard.}\BibitemShut {Stop}%
\bibitem [{\citenamefont {Tikkanen}\ \emph {et~al.}(2020)\citenamefont {Tikkanen}, \citenamefont {Gunja}, \citenamefont {FitzGerald},\ and\ \citenamefont {Zephyrin}}]{noauthor_maternal_2020}%
  \BibitemOpen
  \bibfield  {author} {\bibinfo {author} {\bibfnamefont {R.}~\bibnamefont {Tikkanen}}, \bibinfo {author} {\bibfnamefont {M.~Z.}\ \bibnamefont {Gunja}}, \bibinfo {author} {\bibfnamefont {M.}~\bibnamefont {FitzGerald}},\ and\ \bibinfo {author} {\bibfnamefont {L.~C.}\ \bibnamefont {Zephyrin}},\ }\href {https://doi.org/10.26099/411v-9255} {\bibinfo {title} {Maternal {Mortality} and {Maternity} {Care} in the {United} {States} {Compared} to 10 {Other} {Developed} {Countries}}} (\bibinfo {year} {2020})\BibitemShut {NoStop}%
\bibitem [{\citenamefont {Hoyert}(2024)}]{hoyert_maternal_2024}%
  \BibitemOpen
  \bibfield  {author} {\bibinfo {author} {\bibfnamefont {D.~L.}\ \bibnamefont {Hoyert}},\ }\bibfield  {title} {\bibinfo {title} {Maternal {Mortality} {Rates} in the {United} {States} 2022},\ }\href {https://dx.doi.org/10.15620/cdc/152992} {\bibfield  {journal} {\bibinfo  {journal} {NCHS Health E-Stats}\ } (\bibinfo {year} {2024})}\BibitemShut {NoStop}%
\bibitem [{\citenamefont {Speich}\ \emph {et~al.}(2018)\citenamefont {Speich}, \citenamefont {von Niederhäusern}, \citenamefont {Schur}, \citenamefont {Hemkens}, \citenamefont {Fürst}, \citenamefont {Bhatnagar}, \citenamefont {Alturki}, \citenamefont {Agarwal}, \citenamefont {Kasenda}, \citenamefont {Pauli-Magnus}, \citenamefont {Schwenkglenks},\ and\ \citenamefont {Briel}}]{speich_systematic_2018}%
  \BibitemOpen
  \bibfield  {author} {\bibinfo {author} {\bibfnamefont {B.}~\bibnamefont {Speich}}, \bibinfo {author} {\bibfnamefont {B.}~\bibnamefont {von Niederhäusern}}, \bibinfo {author} {\bibfnamefont {N.}~\bibnamefont {Schur}}, \bibinfo {author} {\bibfnamefont {L.~G.}\ \bibnamefont {Hemkens}}, \bibinfo {author} {\bibfnamefont {T.}~\bibnamefont {Fürst}}, \bibinfo {author} {\bibfnamefont {N.}~\bibnamefont {Bhatnagar}}, \bibinfo {author} {\bibfnamefont {R.}~\bibnamefont {Alturki}}, \bibinfo {author} {\bibfnamefont {A.}~\bibnamefont {Agarwal}}, \bibinfo {author} {\bibfnamefont {B.}~\bibnamefont {Kasenda}}, \bibinfo {author} {\bibfnamefont {C.}~\bibnamefont {Pauli-Magnus}}, \bibinfo {author} {\bibfnamefont {M.}~\bibnamefont {Schwenkglenks}},\ and\ \bibinfo {author} {\bibfnamefont {M.}~\bibnamefont {Briel}},\ }\bibfield  {title} {\bibinfo {title} {Systematic review on costs and resource use of randomized clinical trials shows a lack of transparent and comprehensive data},\ }\href
  {https://doi.org/10.1016/j.jclinepi.2017.12.018} {\bibfield  {journal} {\bibinfo  {journal} {Journal of Clinical Epidemiology}\ }\textbf {\bibinfo {volume} {96}},\ \bibinfo {pages} {1} (\bibinfo {year} {2018})}\BibitemShut {NoStop}%
\bibitem [{\citenamefont {Liu}\ and\ \citenamefont {Mager}(2016)}]{liu_womens_2016}%
  \BibitemOpen
  \bibfield  {author} {\bibinfo {author} {\bibfnamefont {K.~A.}\ \bibnamefont {Liu}}\ and\ \bibinfo {author} {\bibfnamefont {N.~A.~D.}\ \bibnamefont {Mager}},\ }\bibfield  {title} {\bibinfo {title} {Women’s involvement in clinical trials: historical perspective and future implications},\ }\href {https://doi.org/10.18549/PharmPract.2016.01.708} {\bibfield  {journal} {\bibinfo  {journal} {Pharmacy Practice}\ }\textbf {\bibinfo {volume} {14}},\ \bibinfo {pages} {708} (\bibinfo {year} {2016})}\BibitemShut {NoStop}%
\bibitem [{\citenamefont {Dolley}(2018)}]{dolley_big_2018}%
  \BibitemOpen
  \bibfield  {author} {\bibinfo {author} {\bibfnamefont {S.}~\bibnamefont {Dolley}},\ }\bibfield  {title} {\bibinfo {title} {Big {Data}’s {Role} in {Precision} {Public} {Health}},\ }\href {https://doi.org/10.3389/fpubh.2018.00068} {\bibfield  {journal} {\bibinfo  {journal} {Frontiers in Public Health}\ }\textbf {\bibinfo {volume} {6}},\ \bibinfo {pages} {68} (\bibinfo {year} {2018})}\BibitemShut {NoStop}%
\bibitem [{\citenamefont {Salathé}(2018)}]{salathe_digital_2018}%
  \BibitemOpen
  \bibfield  {author} {\bibinfo {author} {\bibfnamefont {M.}~\bibnamefont {Salathé}},\ }\bibfield  {title} {\bibinfo {title} {Digital epidemiology: what is it, and where is it going?},\ }\href {https://doi.org/10.1186/s40504-017-0065-7} {\bibfield  {journal} {\bibinfo  {journal} {Life Sciences, Society and Policy}\ }\textbf {\bibinfo {volume} {14}},\ \bibinfo {pages} {1} (\bibinfo {year} {2018})}\BibitemShut {NoStop}%
\bibitem [{\citenamefont {Stupinski}\ \emph {et~al.}(2022)\citenamefont {Stupinski}, \citenamefont {Alshaabi}, \citenamefont {Arnold}, \citenamefont {Adams}, \citenamefont {Minot}, \citenamefont {Price}, \citenamefont {Dodds},\ and\ \citenamefont {Danforth}}]{stupinski_quantifying_2022}%
  \BibitemOpen
  \bibfield  {author} {\bibinfo {author} {\bibfnamefont {A.~M.}\ \bibnamefont {Stupinski}}, \bibinfo {author} {\bibfnamefont {T.}~\bibnamefont {Alshaabi}}, \bibinfo {author} {\bibfnamefont {M.~V.}\ \bibnamefont {Arnold}}, \bibinfo {author} {\bibfnamefont {J.~L.}\ \bibnamefont {Adams}}, \bibinfo {author} {\bibfnamefont {J.~R.}\ \bibnamefont {Minot}}, \bibinfo {author} {\bibfnamefont {M.}~\bibnamefont {Price}}, \bibinfo {author} {\bibfnamefont {P.~S.}\ \bibnamefont {Dodds}},\ and\ \bibinfo {author} {\bibfnamefont {C.~M.}\ \bibnamefont {Danforth}},\ }\bibfield  {title} {\bibinfo {title} {Quantifying {Changes} in the {Language} {Used} {Around} {Mental} {Health} on {Twitter} {Over} 10 {Years}: {Observational} {Study}},\ }\href {https://doi.org/10.2196/33685} {\bibfield  {journal} {\bibinfo  {journal} {JMIR Mental Health}\ }\textbf {\bibinfo {volume} {9}},\ \bibinfo {pages} {e33685} (\bibinfo {year} {2022})},\ \bibinfo {note} {company: JMIR Mental Health Distributor: JMIR Mental Health Institution: JMIR Mental
  Health Label: JMIR Mental Health Publisher: JMIR Publications Inc., Toronto, Canada}\BibitemShut {NoStop}%
\bibitem [{\citenamefont {Tušl}\ \emph {et~al.}(2022)\citenamefont {Tušl}, \citenamefont {Thelen}, \citenamefont {Marcus}, \citenamefont {Peters}, \citenamefont {Shalaeva}, \citenamefont {Scheckel}, \citenamefont {Sykora}, \citenamefont {Elayan}, \citenamefont {Naslund}, \citenamefont {Shankardass}, \citenamefont {Mooney}, \citenamefont {Fadda},\ and\ \citenamefont {Gruebner}}]{tusl_opportunities_2022}%
  \BibitemOpen
  \bibfield  {author} {\bibinfo {author} {\bibfnamefont {M.}~\bibnamefont {Tušl}}, \bibinfo {author} {\bibfnamefont {A.}~\bibnamefont {Thelen}}, \bibinfo {author} {\bibfnamefont {K.}~\bibnamefont {Marcus}}, \bibinfo {author} {\bibfnamefont {A.}~\bibnamefont {Peters}}, \bibinfo {author} {\bibfnamefont {E.}~\bibnamefont {Shalaeva}}, \bibinfo {author} {\bibfnamefont {B.}~\bibnamefont {Scheckel}}, \bibinfo {author} {\bibfnamefont {M.}~\bibnamefont {Sykora}}, \bibinfo {author} {\bibfnamefont {S.}~\bibnamefont {Elayan}}, \bibinfo {author} {\bibfnamefont {J.~A.}\ \bibnamefont {Naslund}}, \bibinfo {author} {\bibfnamefont {K.}~\bibnamefont {Shankardass}}, \bibinfo {author} {\bibfnamefont {S.~J.}\ \bibnamefont {Mooney}}, \bibinfo {author} {\bibfnamefont {M.}~\bibnamefont {Fadda}},\ and\ \bibinfo {author} {\bibfnamefont {O.}~\bibnamefont {Gruebner}},\ }\bibfield  {title} {\bibinfo {title} {Opportunities and challenges of using social media big data to assess mental health consequences of the {COVID}-19 crisis and
  future major events},\ }\href {https://doi.org/10.1007/s44192-022-00017-y} {\bibfield  {journal} {\bibinfo  {journal} {Discover Mental Health}\ }\textbf {\bibinfo {volume} {2}},\ \bibinfo {pages} {14} (\bibinfo {year} {2022})}\BibitemShut {NoStop}%
\bibitem [{\citenamefont {Grossi}(2025)}]{noauthor_hhs_2025}%
  \BibitemOpen
  \bibfield  {author} {\bibinfo {author} {\bibfnamefont {G.}~\bibnamefont {Grossi}},\ }\href {https://www.ajmc.com/view/hhs-cuts-funding-for-nih-based-women-s-health-initiative-threatening-decades-long-study} {\bibinfo {title} {{HHS} {Cuts} {Funding} for {NIH}-{Based} {Women}'s {Health} {Initiative} {Threatening} {Decades}-{Long} {Study}}} (\bibinfo {year} {2025})\BibitemShut {NoStop}%
\bibitem [{\citenamefont {Kallen}\ \emph {et~al.}(2025)\citenamefont {Kallen}, \citenamefont {Whirledge}, \citenamefont {Goldman},\ and\ \citenamefont {Johnson}}]{kallen_undermining_2025}%
  \BibitemOpen
  \bibfield  {author} {\bibinfo {author} {\bibfnamefont {A.~N.}\ \bibnamefont {Kallen}}, \bibinfo {author} {\bibfnamefont {S.}~\bibnamefont {Whirledge}}, \bibinfo {author} {\bibfnamefont {K.~N.}\ \bibnamefont {Goldman}},\ and\ \bibinfo {author} {\bibfnamefont {J.}~\bibnamefont {Johnson}},\ }\bibfield  {title} {\bibinfo {title} {Undermining {Women}’s {Health} {Research} — {Gambling} with the {Public}’s {Health}},\ }\href {https://doi.org/10.1056/NEJMp2503576} {\bibfield  {journal} {\bibinfo  {journal} {New England Journal of Medicine}\ }\textbf {\bibinfo {volume} {392}},\ \bibinfo {pages} {2185} (\bibinfo {year} {2025})},\ \bibinfo {note} {publisher: Massachusetts Medical Society \_eprint: https://www.nejm.org/doi/pdf/10.1056/NEJMp2503576}\BibitemShut {NoStop}%
\bibitem [{\citenamefont {Sepahpour-Fard}\ and\ \citenamefont {Quayle}(2022)}]{sepahpourfard2022parenting}%
  \BibitemOpen
  \bibfield  {author} {\bibinfo {author} {\bibfnamefont {M.}~\bibnamefont {Sepahpour-Fard}}\ and\ \bibinfo {author} {\bibfnamefont {M.}~\bibnamefont {Quayle}},\ }\bibfield  {title} {\bibinfo {title} {How do mothers and fathers talk about parenting to different audiences?: Stereotypes and audience effects: An analysis of r/daddit, r/mommit, and r/parenting using topic modelling},\ }in\ \href {https://doi.org/10.1145/3485447.3512138} {\emph {\bibinfo {booktitle} {Proceedings of the ACM Web Conference 2022 (WWW '22)}}}\ (\bibinfo  {publisher} {ACM},\ \bibinfo {address} {Virtual Event, Lyon, France},\ \bibinfo {year} {2022})\ p.~\bibinfo {pages} {11},\ \bibinfo {note} {april 25--29, 2022}\BibitemShut {NoStop}%
\bibitem [{\citenamefont {Locke}(2021)}]{locke2021bookreview}%
  \BibitemOpen
  \bibfield  {author} {\bibinfo {author} {\bibfnamefont {A.}~\bibnamefont {Locke}},\ }\bibfield  {title} {\bibinfo {title} {Book review: Language, gender and parenthood online: Negotiating motherhood in mumsnet talk by jai mackenzie},\ }\href {https://doi.org/10.1177/09593535211011304} {\bibfield  {journal} {\bibinfo  {journal} {Feminism \& Psychology}\ }\textbf {\bibinfo {volume} {32}},\ \bibinfo {pages} {125} (\bibinfo {year} {2021})}\BibitemShut {NoStop}%
\bibitem [{\citenamefont {Ali}(2018)}]{ali2018womens}%
  \BibitemOpen
  \bibfield  {author} {\bibinfo {author} {\bibfnamefont {E.}~\bibnamefont {Ali}},\ }\bibfield  {title} {\bibinfo {title} {Women’s experiences with postpartum anxiety disorders: A narrative literature review},\ }\href {https://doi.org/10.2147/IJWH.S158621} {\bibfield  {journal} {\bibinfo  {journal} {International Journal of Women’s Health}\ }\textbf {\bibinfo {volume} {10}},\ \bibinfo {pages} {237} (\bibinfo {year} {2018})}\BibitemShut {NoStop}%
\bibitem [{\citenamefont {Davies}\ \emph {et~al.}(2022)\citenamefont {Davies}, \citenamefont {Todd-Leonida}, \citenamefont {Fallon},\ and\ \citenamefont {Silverio}}]{davies_exclusive_2022}%
  \BibitemOpen
  \bibfield  {author} {\bibinfo {author} {\bibfnamefont {S.~M.}\ \bibnamefont {Davies}}, \bibinfo {author} {\bibfnamefont {B.~F.}\ \bibnamefont {Todd-Leonida}}, \bibinfo {author} {\bibfnamefont {V.~M.}\ \bibnamefont {Fallon}},\ and\ \bibinfo {author} {\bibfnamefont {S.~A.}\ \bibnamefont {Silverio}},\ }\bibfield  {title} {\bibinfo {title} {Exclusive {Breastfeeding} {Duration} and {Perceptions} of {Infant} {Sleep}: {The} {Mediating} {Role} of {Postpartum} {Anxiety}},\ }\href {https://doi.org/10.3390/ijerph19084494} {\bibfield  {journal} {\bibinfo  {journal} {International Journal of Environmental Research and Public Health}\ }\textbf {\bibinfo {volume} {19}},\ \bibinfo {pages} {4494} (\bibinfo {year} {2022})}\BibitemShut {NoStop}%
\bibitem [{\citenamefont {Jackson}\ \emph {et~al.}(2021)\citenamefont {Jackson}, \citenamefont {De~Pascalis}, \citenamefont {Harrold},\ and\ \citenamefont {Fallon}}]{jackson_guilt_2021}%
  \BibitemOpen
  \bibfield  {author} {\bibinfo {author} {\bibfnamefont {L.}~\bibnamefont {Jackson}}, \bibinfo {author} {\bibfnamefont {L.}~\bibnamefont {De~Pascalis}}, \bibinfo {author} {\bibfnamefont {J.}~\bibnamefont {Harrold}},\ and\ \bibinfo {author} {\bibfnamefont {V.}~\bibnamefont {Fallon}},\ }\bibfield  {title} {\bibinfo {title} {Guilt, shame, and postpartum infant feeding outcomes: {A} systematic review},\ }\href {https://doi.org/10.1111/mcn.13141} {\bibfield  {journal} {\bibinfo  {journal} {Maternal \& Child Nutrition}\ }\textbf {\bibinfo {volume} {17}},\ \bibinfo {pages} {e13141} (\bibinfo {year} {2021})}\BibitemShut {NoStop}%
\bibitem [{\citenamefont {Russell}\ \emph {et~al.}(2021)\citenamefont {Russell}, \citenamefont {Birtel}, \citenamefont {Smith}, \citenamefont {Hart},\ and\ \citenamefont {Newman}}]{russell_infant_2021}%
  \BibitemOpen
  \bibfield  {author} {\bibinfo {author} {\bibfnamefont {P.~S.}\ \bibnamefont {Russell}}, \bibinfo {author} {\bibfnamefont {M.~D.}\ \bibnamefont {Birtel}}, \bibinfo {author} {\bibfnamefont {D.~M.}\ \bibnamefont {Smith}}, \bibinfo {author} {\bibfnamefont {K.}~\bibnamefont {Hart}},\ and\ \bibinfo {author} {\bibfnamefont {R.}~\bibnamefont {Newman}},\ }\bibfield  {title} {\bibinfo {title} {Infant feeding and internalized stigma: {The} role of guilt and shame},\ }\href {https://doi.org/10.1111/jasp.12810} {\bibfield  {journal} {\bibinfo  {journal} {Journal of Applied Social Psychology}\ }\textbf {\bibinfo {volume} {51}},\ \bibinfo {pages} {906} (\bibinfo {year} {2021})}\BibitemShut {NoStop}%
\bibitem [{Note2()}]{Note2}%
  \BibitemOpen
  \bibinfo {note} {About one in three women will meet the criteria for anxiety disorders during their lifetime, versus roughly one in five men~\cite {mclean2011gender}.}\BibitemShut {Stop}%
\bibitem [{\citenamefont {Grootendorst}(2022)}]{grootendorst_bertopic_2022}%
  \BibitemOpen
  \bibfield  {author} {\bibinfo {author} {\bibfnamefont {M.}~\bibnamefont {Grootendorst}},\ }\href {https://doi.org/10.48550/arXiv.2203.05794} {\bibinfo {title} {{BERTopic}: {Neural} topic modeling with a class-based {TF}-{IDF} procedure}} (\bibinfo {year} {2022}),\ \bibinfo {note} {arXiv:2203.05794 [cs]}\BibitemShut {NoStop}%
\bibitem [{\citenamefont {Dodds}\ \emph {et~al.}(2023)\citenamefont {Dodds}, \citenamefont {Minot}, \citenamefont {Arnold}, \citenamefont {Alshaabi}, \citenamefont {Adams}, \citenamefont {Dewhurst}, \citenamefont {Gray}, \citenamefont {Frank}, \citenamefont {Reagan},\ and\ \citenamefont {Danforth}}]{dodds_allotaxonometry_2023}%
  \BibitemOpen
  \bibfield  {author} {\bibinfo {author} {\bibfnamefont {P.~S.}\ \bibnamefont {Dodds}}, \bibinfo {author} {\bibfnamefont {J.~R.}\ \bibnamefont {Minot}}, \bibinfo {author} {\bibfnamefont {M.~V.}\ \bibnamefont {Arnold}}, \bibinfo {author} {\bibfnamefont {T.}~\bibnamefont {Alshaabi}}, \bibinfo {author} {\bibfnamefont {J.~L.}\ \bibnamefont {Adams}}, \bibinfo {author} {\bibfnamefont {D.~R.}\ \bibnamefont {Dewhurst}}, \bibinfo {author} {\bibfnamefont {T.~J.}\ \bibnamefont {Gray}}, \bibinfo {author} {\bibfnamefont {M.~R.}\ \bibnamefont {Frank}}, \bibinfo {author} {\bibfnamefont {A.~J.}\ \bibnamefont {Reagan}},\ and\ \bibinfo {author} {\bibfnamefont {C.~M.}\ \bibnamefont {Danforth}},\ }\href {http://arxiv.org/abs/2002.09770} {\bibinfo {title} {Allotaxonometry and rank-turbulence divergence: {A} universal instrument for comparing complex systems}} (\bibinfo {year} {2023}),\ \bibinfo {note} {arXiv:2002.09770 [physics]}\BibitemShut {NoStop}%
\bibitem [{\citenamefont {McInnes}\ \emph {et~al.}(2020)\citenamefont {McInnes}, \citenamefont {Healy},\ and\ \citenamefont {Melville}}]{mcinnes_umap_2020}%
  \BibitemOpen
  \bibfield  {author} {\bibinfo {author} {\bibfnamefont {L.}~\bibnamefont {McInnes}}, \bibinfo {author} {\bibfnamefont {J.}~\bibnamefont {Healy}},\ and\ \bibinfo {author} {\bibfnamefont {J.}~\bibnamefont {Melville}},\ }\bibfield  {title} {\bibinfo {title} {{UMAP}: {Uniform} {Manifold} {Approximation} and {Projection} for {Dimension} {Reduction}},\ }\href@noop {} {\bibfield  {journal} {\bibinfo  {journal} {arXiv}\ }\textbf {\bibinfo {volume} {1802.03426v3}} (\bibinfo {year} {2020})}\BibitemShut {NoStop}%
\bibitem [{\citenamefont {McInnes}\ \emph {et~al.}(2017)\citenamefont {McInnes}, \citenamefont {Healy},\ and\ \citenamefont {Astels}}]{mcinnes_hdbscan_2017}%
  \BibitemOpen
  \bibfield  {author} {\bibinfo {author} {\bibfnamefont {L.}~\bibnamefont {McInnes}}, \bibinfo {author} {\bibfnamefont {J.}~\bibnamefont {Healy}},\ and\ \bibinfo {author} {\bibfnamefont {S.}~\bibnamefont {Astels}},\ }\bibfield  {title} {\bibinfo {title} {hdbscan: {Hierarchical} density based clustering},\ }\href {https://doi.org/10.21105/joss.00205} {\bibfield  {journal} {\bibinfo  {journal} {The Journal of Open Source Software}\ }\textbf {\bibinfo {volume} {2}},\ \bibinfo {pages} {205} (\bibinfo {year} {2017})}\BibitemShut {NoStop}%
\bibitem [{\citenamefont {George}\ and\ \citenamefont {Sumathy}(2023)}]{george_integrated_2023}%
  \BibitemOpen
  \bibfield  {author} {\bibinfo {author} {\bibfnamefont {L.}~\bibnamefont {George}}\ and\ \bibinfo {author} {\bibfnamefont {P.}~\bibnamefont {Sumathy}},\ }\bibfield  {title} {\bibinfo {title} {An integrated clustering and {BERT} framework for improved topic modeling},\ }\href {https://doi.org/10.1007/s41870-023-01268-w} {\bibfield  {journal} {\bibinfo  {journal} {International Journal of Information Technology}\ }\textbf {\bibinfo {volume} {15}},\ \bibinfo {pages} {2187} (\bibinfo {year} {2023})}\BibitemShut {NoStop}%
\bibitem [{\citenamefont {Kherwa}\ and\ \citenamefont {Bansal}(2020)}]{kherwa2020topic}%
  \BibitemOpen
  \bibfield  {author} {\bibinfo {author} {\bibfnamefont {P.}~\bibnamefont {Kherwa}}\ and\ \bibinfo {author} {\bibfnamefont {P.}~\bibnamefont {Bansal}},\ }\bibfield  {title} {\bibinfo {title} {Topic modeling: A comprehensive review},\ }\href {https://doi.org/10.4108/eai.13-7-2018.159623} {\bibfield  {journal} {\bibinfo  {journal} {EAI Endorsed Transactions on Scalable Information Systems}\ }\textbf {\bibinfo {volume} {7}},\ \bibinfo {pages} {1} (\bibinfo {year} {2020})}\BibitemShut {NoStop}%
\bibitem [{\citenamefont {Egger}\ and\ \citenamefont {Yu}(2022)}]{egger2022topic}%
  \BibitemOpen
  \bibfield  {author} {\bibinfo {author} {\bibfnamefont {R.}~\bibnamefont {Egger}}\ and\ \bibinfo {author} {\bibfnamefont {J.}~\bibnamefont {Yu}},\ }\bibfield  {title} {\bibinfo {title} {Topic modeling comparison between lda, top2vec, and bertopic to demystify twitter posts},\ }\href {https://doi.org/10.3389/fsoc.2022.886498} {\bibfield  {journal} {\bibinfo  {journal} {Frontiers in Sociology}\ }\textbf {\bibinfo {volume} {7}},\ \bibinfo {pages} {1} (\bibinfo {year} {2022})}\BibitemShut {NoStop}%
\bibitem [{\citenamefont {Rogers}\ and\ \citenamefont {Longo}(2017)}]{rogers2017comparison}%
  \BibitemOpen
  \bibfield  {author} {\bibinfo {author} {\bibfnamefont {N.}~\bibnamefont {Rogers}}\ and\ \bibinfo {author} {\bibfnamefont {L.}~\bibnamefont {Longo}},\ }\bibfield  {title} {\bibinfo {title} {A comparison on the classification of short-text documents using latent dirichlet allocation and formal concept analysis},\ }in\ \href@noop {} {\emph {\bibinfo {booktitle} {CEUR Workshop Proceedings}}},\ Vol.\ \bibinfo {volume} {2086}\ (\bibinfo {year} {2017})\ pp.\ \bibinfo {pages} {50--62}\BibitemShut {NoStop}%
\bibitem [{\citenamefont {Alkhafaji}\ and\ \citenamefont {Al-Rashid}(2021)}]{alkhafaji2021topic}%
  \BibitemOpen
  \bibfield  {author} {\bibinfo {author} {\bibfnamefont {D.~W.}\ \bibnamefont {Alkhafaji}}\ and\ \bibinfo {author} {\bibfnamefont {S.~A.}\ \bibnamefont {Al-Rashid}},\ }\bibfield  {title} {\bibinfo {title} {Topic modeling for clustering arabic documents},\ }in\ \href {https://doi.org/10.1109/IT-ELA52201.2021.9773538} {\emph {\bibinfo {booktitle} {Proceedings of the 2nd Information Technology to Enhance E-Learning and Other Applications Conference (IT-ELA 2021)}}}\ (\bibinfo {year} {2021})\ pp.\ \bibinfo {pages} {76--81}\BibitemShut {NoStop}%
\bibitem [{\citenamefont {Rahman}\ \emph {et~al.}(2020)\citenamefont {Rahman}, \citenamefont {Prama},\ and\ \citenamefont {Anwar}}]{Rahman2020ModelingTS}%
  \BibitemOpen
  \bibfield  {author} {\bibinfo {author} {\bibfnamefont {M.~H.}\ \bibnamefont {Rahman}}, \bibinfo {author} {\bibfnamefont {T.~T.}\ \bibnamefont {Prama}},\ and\ \bibinfo {author} {\bibfnamefont {M.~M.}\ \bibnamefont {Anwar}},\ }\bibfield  {title} {\bibinfo {title} {Modeling topic specific credibility in twitter based on structural and attribute properties},\ }in\ \href {https://api.semanticscholar.org/CorpusID:233329338} {\emph {\bibinfo {booktitle} {International Conference on Health Information Science}}}\ (\bibinfo {year} {2020})\BibitemShut {NoStop}%
\bibitem [{\citenamefont {Benites-Lazaro}\ \emph {et~al.}(2018)\citenamefont {Benites-Lazaro}, \citenamefont {Giatti},\ and\ \citenamefont {Giarolla}}]{benites-lazaro_topic_2018}%
  \BibitemOpen
  \bibfield  {author} {\bibinfo {author} {\bibfnamefont {L.}~\bibnamefont {Benites-Lazaro}}, \bibinfo {author} {\bibfnamefont {L.}~\bibnamefont {Giatti}},\ and\ \bibinfo {author} {\bibfnamefont {A.}~\bibnamefont {Giarolla}},\ }\bibfield  {title} {\bibinfo {title} {Topic modeling method for analyzing social actor discourses on climate change, energy and food security},\ }\href {https://doi.org/https://doi.org/10.1016/j.erss.2018.07.031} {\bibfield  {journal} {\bibinfo  {journal} {Energy Research \& Social Science}\ }\textbf {\bibinfo {volume} {45}},\ \bibinfo {pages} {318} (\bibinfo {year} {2018})}\BibitemShut {NoStop}%
\bibitem [{\citenamefont {Blei}\ \emph {et~al.}(2003)\citenamefont {Blei}, \citenamefont {Ng},\ and\ \citenamefont {Jordan}}]{latent_dirichlet}%
  \BibitemOpen
  \bibfield  {author} {\bibinfo {author} {\bibfnamefont {D.~M.}\ \bibnamefont {Blei}}, \bibinfo {author} {\bibfnamefont {A.~Y.}\ \bibnamefont {Ng}},\ and\ \bibinfo {author} {\bibfnamefont {M.~I.}\ \bibnamefont {Jordan}},\ }\bibfield  {title} {\bibinfo {title} {Latent dirichlet allocation},\ }\href {https://dl.acm.org/doi/10.5555/944919.944937} {\bibfield  {journal} {\bibinfo  {journal} {Latent dirichlet allocation {\textbar} {The} {Journal} of {Machine} {Learning} {Research}}\ } (\bibinfo {year} {2003})}\BibitemShut {NoStop}%
\bibitem [{\citenamefont {{American Psychiatric Association}}(2022)}]{apa_dsm5tr_2022}%
  \BibitemOpen
  \bibfield  {author} {\bibinfo {author} {\bibnamefont {{American Psychiatric Association}}},\ }\href@noop {} {\emph {\bibinfo {title} {{Diagnostic and Statistical Manual of Mental Disorders, Fifth Edition, Text Revision (DSM‑5‑TR)}}}}\ (\bibinfo  {publisher} {American Psychiatric Association},\ \bibinfo {address} {Washington, DC},\ \bibinfo {year} {2022})\ \bibinfo {note} {text revision of DSM‑5, includes updated diagnostic criteria, ICD‑10‑CM codes, and new disorders such as prolonged grief disorder}\BibitemShut {NoStop}%
\bibitem [{\citenamefont {Ystrom}(2012)}]{ystrom_breastfeeding_2012}%
  \BibitemOpen
  \bibfield  {author} {\bibinfo {author} {\bibfnamefont {E.}~\bibnamefont {Ystrom}},\ }\bibfield  {title} {\bibinfo {title} {Breastfeeding cessation and symptoms of anxiety and depression: a longitudinal cohort study},\ }\href {https://doi.org/10.1186/1471-2393-12-36} {\bibfield  {journal} {\bibinfo  {journal} {BMC Pregnancy and Childbirth}\ }\textbf {\bibinfo {volume} {12}},\ \bibinfo {pages} {36} (\bibinfo {year} {2012})}\BibitemShut {NoStop}%
\bibitem [{\citenamefont {Coo}\ \emph {et~al.}(2020)\citenamefont {Coo}, \citenamefont {García}, \citenamefont {Mira},\ and\ \citenamefont {Valdés}}]{coo_role_2020}%
  \BibitemOpen
  \bibfield  {author} {\bibinfo {author} {\bibfnamefont {S.}~\bibnamefont {Coo}}, \bibinfo {author} {\bibfnamefont {M.~I.}\ \bibnamefont {García}}, \bibinfo {author} {\bibfnamefont {A.}~\bibnamefont {Mira}},\ and\ \bibinfo {author} {\bibfnamefont {V.}~\bibnamefont {Valdés}},\ }\bibfield  {title} {\bibinfo {title} {The {Role} of {Perinatal} {Anxiety} and {Depression} in {Breastfeeding} {Practices}},\ }\href {https://doi.org/10.1089/bfm.2020.0091} {\bibfield  {journal} {\bibinfo  {journal} {Breastfeeding Medicine}\ }\textbf {\bibinfo {volume} {15}},\ \bibinfo {pages} {495} (\bibinfo {year} {2020})},\ \bibinfo {note} {publisher: Mary Ann Liebert, Inc., publishers}\BibitemShut {NoStop}%
\bibitem [{\citenamefont {Nagel}\ \emph {et~al.}(2022)\citenamefont {Nagel}, \citenamefont {Howland}, \citenamefont {Pando}, \citenamefont {Stang}, \citenamefont {Mason}, \citenamefont {Fields},\ and\ \citenamefont {Demerath}}]{nagel_maternal_2022}%
  \BibitemOpen
  \bibfield  {author} {\bibinfo {author} {\bibfnamefont {E.~M.}\ \bibnamefont {Nagel}}, \bibinfo {author} {\bibfnamefont {M.~A.}\ \bibnamefont {Howland}}, \bibinfo {author} {\bibfnamefont {C.}~\bibnamefont {Pando}}, \bibinfo {author} {\bibfnamefont {J.}~\bibnamefont {Stang}}, \bibinfo {author} {\bibfnamefont {S.~M.}\ \bibnamefont {Mason}}, \bibinfo {author} {\bibfnamefont {D.~A.}\ \bibnamefont {Fields}},\ and\ \bibinfo {author} {\bibfnamefont {E.~W.}\ \bibnamefont {Demerath}},\ }\bibfield  {title} {\bibinfo {title} {Maternal psychological distress and lactation and breastfeeding outcomes: {A} narrative review},\ }\href {https://doi.org/10.1016/j.clinthera.2021.11.007} {\bibfield  {journal} {\bibinfo  {journal} {Clinical therapeutics}\ }\textbf {\bibinfo {volume} {44}},\ \bibinfo {pages} {215} (\bibinfo {year} {2022})}\BibitemShut {NoStop}%
\bibitem [{\citenamefont {Thomson}\ \emph {et~al.}(2014)\citenamefont {Thomson}, \citenamefont {Ebisch‐Burton},\ and\ \citenamefont {Flacking}}]{thomson_shame_2014}%
  \BibitemOpen
  \bibfield  {author} {\bibinfo {author} {\bibfnamefont {G.}~\bibnamefont {Thomson}}, \bibinfo {author} {\bibfnamefont {K.}~\bibnamefont {Ebisch‐Burton}},\ and\ \bibinfo {author} {\bibfnamefont {R.}~\bibnamefont {Flacking}},\ }\bibfield  {title} {\bibinfo {title} {Shame if you do – shame if you don't: women's experiences of infant feeding},\ }\href {https://doi.org/10.1111/mcn.12148} {\bibfield  {journal} {\bibinfo  {journal} {Maternal \& Child Nutrition}\ }\textbf {\bibinfo {volume} {11}},\ \bibinfo {pages} {33} (\bibinfo {year} {2014})}\BibitemShut {NoStop}%
\bibitem [{\citenamefont {Scott}\ \emph {et~al.}(2021)\citenamefont {Scott}, \citenamefont {Webb}, \citenamefont {Martyn-St~James}, \citenamefont {Rowse},\ and\ \citenamefont {Weich}}]{scott_improving_2021}%
  \BibitemOpen
  \bibfield  {author} {\bibinfo {author} {\bibfnamefont {A.~J.}\ \bibnamefont {Scott}}, \bibinfo {author} {\bibfnamefont {T.~L.}\ \bibnamefont {Webb}}, \bibinfo {author} {\bibfnamefont {M.}~\bibnamefont {Martyn-St~James}}, \bibinfo {author} {\bibfnamefont {G.}~\bibnamefont {Rowse}},\ and\ \bibinfo {author} {\bibfnamefont {S.}~\bibnamefont {Weich}},\ }\bibfield  {title} {\bibinfo {title} {Improving sleep quality leads to better mental health: {A} meta-analysis of randomised controlled trials},\ }\href {https://doi.org/10.1016/j.smrv.2021.101556} {\bibfield  {journal} {\bibinfo  {journal} {Sleep Medicine Reviews}\ }\textbf {\bibinfo {volume} {60}},\ \bibinfo {pages} {101556} (\bibinfo {year} {2021})}\BibitemShut {NoStop}%
\bibitem [{\citenamefont {Bai}\ and\ \citenamefont {Wunderlich}(2013)}]{bai_lactation_2013}%
  \BibitemOpen
  \bibfield  {author} {\bibinfo {author} {\bibfnamefont {Y.}~\bibnamefont {Bai}}\ and\ \bibinfo {author} {\bibfnamefont {S.~M.}\ \bibnamefont {Wunderlich}},\ }\bibfield  {title} {\bibinfo {title} {Lactation {Accommodation} in the {Workplace} and {Duration} of {Exclusive} {Breastfeeding}},\ }\href {https://doi.org/10.1111/jmwh.12072} {\bibfield  {journal} {\bibinfo  {journal} {Journal of Midwifery \& Women's Health}\ }\textbf {\bibinfo {volume} {58}},\ \bibinfo {pages} {690} (\bibinfo {year} {2013})},\ \bibinfo {note} {\_eprint: https://onlinelibrary.wiley.com/doi/pdf/10.1111/jmwh.12072}\BibitemShut {NoStop}%
\bibitem [{\citenamefont {Magner}\ and\ \citenamefont {Phillipi}(2015)}]{magner_using_2015}%
  \BibitemOpen
  \bibfield  {author} {\bibinfo {author} {\bibfnamefont {A.}~\bibnamefont {Magner}}\ and\ \bibinfo {author} {\bibfnamefont {C.~A.}\ \bibnamefont {Phillipi}},\ }\bibfield  {title} {\bibinfo {title} {Using a {Wellness} {Program} to {Promote} a {Culture} of {Breastfeeding} in the {Workplace}: {Oregon} {Health} \& {Science} {University}’s {Experience}},\ }\href {https://doi.org/10.1177/0890334414554262} {\bibfield  {journal} {\bibinfo  {journal} {Journal of Human Lactation}\ }\textbf {\bibinfo {volume} {31}},\ \bibinfo {pages} {40} (\bibinfo {year} {2015})},\ \bibinfo {note} {publisher: SAGE Publications Inc STM}\BibitemShut {NoStop}%
\bibitem [{\citenamefont {Tsai}(2013)}]{tsai_impact_2013}%
  \BibitemOpen
  \bibfield  {author} {\bibinfo {author} {\bibfnamefont {S.-Y.}\ \bibnamefont {Tsai}},\ }\bibfield  {title} {\bibinfo {title} {Impact of a breastfeeding-friendly workplace on an employed mother's intention to continue breastfeeding after returning to work},\ }\href {https://doi.org/10.1089/bfm.2012.0119} {\bibfield  {journal} {\bibinfo  {journal} {Breastfeeding Medicine: The Official Journal of the Academy of Breastfeeding Medicine}\ }\textbf {\bibinfo {volume} {8}},\ \bibinfo {pages} {210} (\bibinfo {year} {2013})}\BibitemShut {NoStop}%
\bibitem [{\citenamefont {Hanna}\ \emph {et~al.}(2005)\citenamefont {Hanna}, \citenamefont {Weinberg}, \citenamefont {Dant},\ and\ \citenamefont {Berger}}]{hanna_internet-based_2005}%
  \BibitemOpen
  \bibfield  {author} {\bibinfo {author} {\bibfnamefont {R.~C.}\ \bibnamefont {Hanna}}, \bibinfo {author} {\bibfnamefont {B.}~\bibnamefont {Weinberg}}, \bibinfo {author} {\bibfnamefont {R.~P.}\ \bibnamefont {Dant}},\ and\ \bibinfo {author} {\bibfnamefont {P.~D.}\ \bibnamefont {Berger}},\ }\bibfield  {title} {\bibinfo {title} {Do internet-based surveys increase personal self-disclosure?},\ }\href {https://doi.org/10.1057/palgrave.dbm.3240270} {\bibfield  {journal} {\bibinfo  {journal} {Journal of Database Marketing \& Customer Strategy Management}\ }\textbf {\bibinfo {volume} {12}},\ \bibinfo {pages} {342} (\bibinfo {year} {2005})}\BibitemShut {NoStop}%
\bibitem [{\citenamefont {Andrade}(2020)}]{andrade_limitations_2020}%
  \BibitemOpen
  \bibfield  {author} {\bibinfo {author} {\bibfnamefont {C.}~\bibnamefont {Andrade}},\ }\bibfield  {title} {\bibinfo {title} {The {Limitations} of {Online} {Surveys}},\ }\href {https://doi.org/10.1177/0253717620957496} {\bibfield  {journal} {\bibinfo  {journal} {Indian Journal of Psychological Medicine}\ }\textbf {\bibinfo {volume} {42}},\ \bibinfo {pages} {575} (\bibinfo {year} {2020})}\BibitemShut {NoStop}%
\bibitem [{\citenamefont {Coughlan}\ \emph {et~al.}(2009)\citenamefont {Coughlan}, \citenamefont {Cronin},\ and\ \citenamefont {Ryan}}]{coughlan_survey_2009}%
  \BibitemOpen
  \bibfield  {author} {\bibinfo {author} {\bibfnamefont {M.}~\bibnamefont {Coughlan}}, \bibinfo {author} {\bibfnamefont {P.}~\bibnamefont {Cronin}},\ and\ \bibinfo {author} {\bibfnamefont {F.}~\bibnamefont {Ryan}},\ }\bibfield  {title} {\bibinfo {title} {Survey research: {Process} and limitations},\ }\href {https://doi.org/10.12968/ijtr.2009.16.1.37935} {\bibfield  {journal} {\bibinfo  {journal} {International Journal of Therapy and Rehabilitation}\ }\textbf {\bibinfo {volume} {16}},\ \bibinfo {pages} {9} (\bibinfo {year} {2009})},\ \bibinfo {note} {publisher: Mark Allen Group}\BibitemShut {NoStop}%
\bibitem [{\citenamefont {Szugye}\ \emph {et~al.}(2023)\citenamefont {Szugye}, \citenamefont {Murra},\ and\ \citenamefont {Lam}}]{szugye_new_2023}%
  \BibitemOpen
  \bibfield  {author} {\bibinfo {author} {\bibfnamefont {H.}~\bibnamefont {Szugye}}, \bibinfo {author} {\bibfnamefont {A.}~\bibnamefont {Murra}},\ and\ \bibinfo {author} {\bibfnamefont {S.~K.}\ \bibnamefont {Lam}},\ }\bibfield  {title} {\bibinfo {title} {A new policy update on breastfeeding: {What} all clinicians need to know},\ }\href {https://doi.org/10.3949/ccjm.90a.22099} {\bibfield  {journal} {\bibinfo  {journal} {Cleveland Clinic Journal of Medicine}\ }\textbf {\bibinfo {volume} {90}},\ \bibinfo {pages} {469} (\bibinfo {year} {2023})},\ \bibinfo {note} {publisher: Cleveland Clinic Journal of Medicine Section: Guidelines to Practice}\BibitemShut {NoStop}%
\end{thebibliography}%

\clearpage
\appendix

\onecolumngrid

\setcounter{page}{1}
\renewcommand{\thepage}{A\arabic{page}}
\renewcommand{\thefigure}{A\arabic{figure}}
\renewcommand{\thetable}{A\arabic{table}}
\setcounter{figure}{0}
\setcounter{table}{0}
\section*{Appendix}

Using \texttt{BERTopic} utilizes Uniform Manifold Approximation and Projection (UMAP) when clustering topics which employs stochastic gradient descent. Thus, different clusters emerge due to randomness introduced in this step. For clarity, we ran \texttt{BERTopic} and use that same mapping throughout our analysis of clusters. However, it is important to note how these clusters can vary based on this step, and a comparison clustering is provided at the end of this section. 

Using the standard \texttt{BERTopic} steps, we report the top 10 topics that emerged in \ref{table:top_20}. We disregard the stop word containing topic as is standard using \texttt{BERTopic}. 

\begin{table}[hbp]
\begin{tabular}{l|c|c}
\bf{Topic} & \bf{Count} &\bf{Name} \\
\hline
1   &       4,226   & 1\_measuring\_percentile\_oz\_lbs \\
2   &      2,726       & 2\_induction\_induced\_induce\_39 \\
3   &       2,604 & 3\_spotting\_bleeding\_brown\_clots \\
4   &       2,260 &   4\_movement\_kicks\_movements\_flutters \\
5   &       1,798 & 5\_sleep\_exhausted\_nap\_tired \\
6   &       1,620 & 6\_dog\_cat\_dogs\_cats \\ 
7   &       1,463 & 7\_gained\_gain\_weight\_pounds \\ 
8   &       1,375 & 8\_section\_recovery\_vaginal\_csection \\ 
9   &       1,330 &    9\_bump\_bloat\_bloated\_showing \\ 
10   &       1,304 &      10\_netflix\_book\_books\_series \\ 
\end{tabular}
\caption{\texttt{BERTopic} identified topics in April 2019 birth club.} \label{table:top_20}
\end{table}
Within the top 10 identified topics, some deal with early- and mid- pregnancy such as early bump/bloating appearance (topic 9) and kicking/movement (topic 4). Other dominant topics (2, 8) deal with labor itself. A few individuals topics are vague and could refer to before and after pregnancy like measuring and percentiles (topic 1), spotting (3), sleep and tiredness (topic 5), and weight gain (topic 7) which occur during pregnancy and post-birth. Topics that do not occur in a logical grouping are topics about cats and dogs (topic 6) with representative documents speaking to clinginess of pets during pregnancy and Netflix/books (topic 10). 

These topics collectively represent the full breadth of concerns across pregnancy and post-birth. Interestingly, breastfeeding does not appear in the broad overview of topics, which may be due to the predominance of other discussions. However, since the Allotaxonometer demonstrated that breastfeeding terms experienced a large shift pre- and post-birth, we then explored breastfeeding specific topics in the data set using the same pipeline previously described. The breastfeeding topics are varied but include conversation of formula feeding versus breastfeeding, sore breasts, and weight loss due to breastfeeding. 

\begin{table}[hbp]
\begin{tabular}{l|c|c}
\bf{Topic} & \bf{Count} &\bf{Name} \\
\hline
15   &       1,141   & 15\_formula\_breastfeeding\_breastfeed\_fed \\
56   &      461       & 56\_sore\_boobs\_breasts\_symptoms \\
196  &       176  & 187\_weight\_breastfeeding\_lose\_lost \\
349 &    94 & 349\_classes\_class\_breastfeeding\_birthing \\
851  &      35 &   851\_lamictal\_breastfeeding\_breastfeed\_lam \\
1004  &       29 & 1004\_pump\_breast\_champva\_willow \\
1052  &       27 & 1052\_aeroflow\_contacted\_insurance... \\ 
1151  &       26 & 1115\_lump\_cancer\_breast\_lumps \\ 
 
\end{tabular}
\caption{\texttt{BERTopic} identified ``breastfeeding'' topics in April 2019 birth club.}
\label{table:top_20}
\end{table}

Initial topic analysis suggested that breastfeeding discussion encompasses more than just feeding concerns and relates to the experience of the breastfeeding person, the impact of breastfeeding on medication usage, and insurance and reimbursement concerns. We then explored the top breastfeeding topic more closely; first, we collected all posts of users who reported struggling breastfeeding in the major breastfeeding topic. Then, we conducted two separate topic modelings; one topic model on all such user posts, and one topic model on all such user posts excluding the specifically identified posts from the breastfeeding topic of the full data set. 

\begin{table}
\begin{tabular}{l|c|c}
\bf{Topic} & \bf{Count} &\bf{Name} \\
\hline
1   &       2,867  & (omitted and/or missing post data) \\
2   &     1,287       & 2\_seat\_stroller\_car\_infant \\
3   &       850 & 3\_percentile\_lbs\_measuring\_oz \\
4   &       824 &   4\_leave\_fmla\_job\_disability \\
5   &       769 & 5\_induction\_induced\_inductions\_induce \\
6   &      717 & 6\_bump\_bloat\_belly\_look \\ 
7   &       712 & 7\_movement\_kicks\_movements\_flutters \\ 
8   &       652 & 8\_sleep\_exhausted\_tired\_nap \\ 
9   &       514 &    9\_formula\_breastfeeding\_breastfeed\_fed \\ 
10   &       637 &      10\_chicken\_potatoes\_sauce\_meals \\ 
\end{tabular}
\caption{\texttt{BERTopic} identified topics in subset of posts of users who self-reported breastfeeding struggles in April 2019 birth club main ``Breastfeeding'' topic.}
\label{table:top_10_inbftopicstruggle}
\end{table}

Our further exploration of the breastfeeding topic including all posts yielded noticeable differences in top 10 identified topics. Breastfeeding, of course, ranked higher than in the full dataset. Terms for induction, sleep, and percentile weights rated about the same as in the full dataset, which may be related to these items effecting many aspects of pregnancy and post-pregnancy experiences. Interestingly, leave and FMLA rated much more highly in the breastfeeding struggling subset of data, which suggests that this may be a correlated risk factor in early cessation of breastfeeding that should be explored further. 

\begin{table}
\begin{tabular}{l|c|c}
\bf{Topic} & \bf{Count} &\bf{Name} \\
\hline
1   &       2,866 & NA (omitted and/or missing post data) \\
2   &     843      & 2\_percentile\_measuring\_lbs\_oz \\
3   &       850 & 3\_induction\_induced\_inductions\_39 \\
4   &       824 &   4\_leave\_fmla\_job\_disability \\
5   &       769 & 5\_vaccine\_vaccines\_flu\_shot \\
6   &      717 & 6\_bump\_bloat\_belly\_look \\ 
7   &       712 & 7\_sleep\_exhausted\_tired\_nap \\ 
8   &       652 & 8\_boy\_girl\_gender\_boys \\ 
9   &       514 &    9\_stroller\_seat\_infant\_car \\ 
10   &       637 &      10\_throat\_cough\_cold\_fever \\ 
\end{tabular}
\caption{\texttt{BERTopic} identified topics in subset of posts of users who self-reported breastfeeding struggles in April 2019 birth club main ``Breastfeeding'' topic, excluding the posts in which they self-reported struggles.}
\label{table:top_10_inbftopicstruggledrop}
\end{table}

In the topic modeling that excluded the specifics posts that indicated self-report of breastfeeding struggle, we notice many of the same trends as the previous topic modeling. Interestingly, a cold/sore throat/fever topic emerges, which may suggest that being ill (whether it be child or parent) could be another risk factor that should be studied further in relation to breastfeeding. 

Additionally, we assessed the relative importance of anxiety and depression labeled topics in the various subsets of data, and noticed that rank and importance of the top anxiety-labeled and depression-labeled topics (that is, in topics that contain the terms anxiety or depression) increased in both subsets of breastfeeding struggles (with and without the coded posts). 

\begin{table}[h]
\begin{tabular}{l|c|c|c}
\bf{Topic tag} & \bf{Subset} & \bf{Full Data} & \bf{Subset without Posts}\\
\hline
Anxiety & 122, 0.0009 & 221, 0.0005 & 117, 0.001 \\
Depression  & 266, 0.0005 & 297, 0.0003 & 198, 0.006 \\

\end{tabular}
\caption{Comparison of rank and percentage of topic in respective corpus for highest ranked anxiety- and depression-labeled topics in subset of self-reported struggling breastfeeding users topics (topic count: 842), full data set (topic count: 2,120), and dropping all identified self-reported struggling breastfeeding posts (topic count: 854).
} \label{table:top_20}
\end{table}

\begin{table}
\begin{tabular}{l|c|c}
\bf{Topic} & \bf{Count} &\bf{Name} \\
\hline
1   &       11,039   & 0\_name\_names\_middle\_named \\
2   &      10,564       & 1\_na\_af\_been\_ \\
3   &       4,224 & 2\_kicks\_movement\_movements\_flutters \\
4   &       4,112 &   3\_measuring\_percentile\_oz\_lbs \\
5   &       2,866 & 4\_bump\_bloat\_showing\_bloated \\
6   &       2,673 & 5\_spotting\_bleeding\_brown\_clots \\ 
7   &       2,495 & 6\_induction\_induced\_induce\_39 \\ 
8   &       1,931 & 7\_sleep\_exhausted\_nap\_tired \\ 
9   &       1,854 &    8\_ultrasound\_ultrasounds\_3d\_elective \\ 
10   &       1,854 &      9\_nausea\_sickness\_nauseous\_sick \\ 
\end{tabular}
\caption{\texttt{BERTopic} identified topics in April 2019 birth club, alternate run.}
\label{table:top_10}
\vspace{1cm}
\begin{tabular}{l|c|c}
\bf{Topic} & \bf{Count} &\bf{Name} \\
\hline
153   &       197   & 153\_breastfeeding\_breastfeed\_breastfed\_wean \\
187   &      171       & 187\_formula\_fed\_breastfeeding\_feeding \\
224  &       143 & 224\_breastfeeding\_weight\_lose\_lost \\
232 & 136 & 232\_poop\_pooping\_poops\_pooped \\
234   &      134 &   234\_class\_classes\_breastfeeding\_birthing \\
331  &       94 & 331\_awesome\_cool\_wow\_thats \\
725   &       41 & 725\_pills\_diet\_breastfeeding\_colace \\ 
800   &       38 & 800\_lamictal\_breastfeeding\_breastfeed\_lam \\ 
1256   &       21 & 1256\_latch\_latching\_formula\_nipple \\ 
1733   &       14 &    1733\_wake\_wakes\_feed\_eats \\ 
2027   &       10 &      2027\_nursing\_patpat\_tops\_blousesshirts \\ 
2028  &       10 &   2028\_nipples\_nipple\_breastfeeding\_painlike \\ 
 
\end{tabular}
\caption{\texttt{BERTopic} identified ``breastfeeding'' topics in April 2019 birth club, alternate run.}
\label{table:top_20}
\end{table}

\begin{table}
\begin{tabular}{l|c}
\bf{Classifier} & \bf{Accuracy} \\
\hline
Logistic Regression & 0.92 \\ 
K-Nearest Neighbors & 0.92 \\
Support Vector Machine & 0.92 \\
MultinomialNB & 0.92 \\
Decision Tree & 0.89 \\
Random Forest & 0.92 \\
Gradient Boosting & 0.88 \\
AdaBoost & 0.91 \\
Perceptron & 0.90 \\
Ridge Classifier & 0.90 \\
Nearest Centroid & 0.70 \\ 
 
\end{tabular}
\caption{Classifiers' accuracy on human labeled data ($n=1,141$ documents, 3 human coders) (80\% train/20\% test). Using three coders (CB, IS, and PS), we independently annotated the predominant breastfeeding topic in the April 2019 birth club to identify self-report of struggles associated with breastfeeding. Our inclusion criteria centered on explicit mention of physical or mental struggles during breastfeeding (e.g. pain, soreness, frustration, worry) as well as self-report of cessation of breastfeeding before 1 year (based on the 2019 American Academy of Pediatricians recommendation of 1 year duration of exclusive breastfeeding, which changed to 6 months as of 2023~\cite{szugye_new_2023}). We excluded anecdotes from users about other people who struggled with breastfeeding (e.g., relating a story about a friend who struggled) and attempt to identify current pregnancy/birth struggles (i.e., non-retrospective about previous births before April 2019). We required all coders to rate a post as ``negative'' to include the breastfeeding struggle in the subset topic modeling; if coders disagreed, they discussed its inclusion until we reached consensus. Upon initial coding, prior to discussion, coders achieved 83.3\% agreement (590/708 posts). }
\label{table:classifiers}
\end{table}

\begin{table}[h!]
\centering
\begin{tabular}{|l|l|c|c|}
\hline
\texttt{doc\_topic\_prior} ($\alpha$) & \texttt{topic\_word\_prior} ($\eta$) & number of documents ($k$) & perplexity score \\
\hline
\texttt{0.25}  & \texttt{0.25}  & 4   & 2261.9289 \\
\texttt{0.1}   & \texttt{0.1}   & 10  & 2189.5068 \\
\texttt{0.05}  & \texttt{0.05}  & 20  & 2135.6854 \\
\texttt{0.02}  & \texttt{0.02}  & 50  & 2276.6576 \\
\texttt{0.01}  & \texttt{0.01}  & 100 & 2588.2222 \\
\texttt{0.005} & \texttt{0.005} & 200 & 3192.3299 \\
\texttt{0.25}  & \texttt{0.1}   & 4   & 2240.2458 \\
\texttt{0.1}   & \texttt{0.1}   & 10  & 2189.5068 \\
\texttt{0.05}  & \texttt{0.1}   & 20  & 2179.0988 \\
\texttt{0.02}  & \texttt{0.1}   & 50  & 2481.3575 \\
\texttt{0.01}  & \texttt{0.1}   & 100 & 3155.0606 \\
\texttt{0.005} & \texttt{0.1}   & 200 & 4751.3159 \\
\texttt{0.01}  & \texttt{0.25}  & 4   & 2279.6107 \\
\texttt{0.01}  & \texttt{0.1}   & 10  & 2235.3667 \\
\texttt{0.01}  & \texttt{0.05}  & 20  & 2208.6584 \\
\texttt{0.01}  & \texttt{0.02}  & 50  & 2366.8750 \\
\texttt{0.01}  & \texttt{0.01}  & 100 & 2588.2222 \\
\texttt{0.01}  & \texttt{0.005} & 200 & 2934.7749 \\
\hline
\end{tabular}
\caption{LDA hyperparameter combinations and resulting perplexity using \texttt{sk-learn} Latent Dirichlet Allocation (LDA) function. The parameter settings that yielded the lowest perplexity score on the full dataset was $\alpha=0.5, \eta=0.5, k=20$.}
\label{parameter_sweep}
\end{table}



\end{document}